\documentclass[final,1p,times]{elsarticle}

\usepackage{amssymb}
\usepackage{amsmath}
\usepackage{bm}
\usepackage{graphicx}

\biboptions{comma,compress}
\newcommand{\bea}{\begin{eqnarray}}
\newcommand{\eea}{\end{eqnarray}}

\journal{Annals of Physics}

\begin{document}

\begin{frontmatter}




\title{Nonperturbative $NN$ scattering in $^3S_1$$-$$^3D_1$
channels of EFT($\not\!\pi$)}


\author[a,b]{Ji-Feng Yang}
\ead{jfyang@phy.ecnu.edu.cn}
\address[a]{Department of Physics, East China Normal University,
Shanghai 200241, China}
\address[b]{KITPC, Chinese Academy of Sciences, Beijing 100190,
China}

\begin{abstract}
The closed-form $T$ matrices in the $^3S_1$$-$$^3D_1$ channels of EFT($\not\!\!\pi$) for $NN$ scattering with the potentials truncated at order $\mathcal{O}(Q^4)$ are presented with the nonperturbative divergences parametrized in a general manner. The stringent constraints imposed by the closed form of the $T$ matrices are exploited in the underlying theory perspective and turned into virtues in the implementation of subtractions and the manifestation of power counting rules in nonperturbative regimes, leading us to the concept of EFT scenario. A number of scenarios of the EFT description of $NN$ scattering are compared with PSA data in terms of effective range expansion and $^3S_1$ phase shifts, showing that it is favorable to proceed in a scenario with conventional EFT couplings and sophisticated renormalization in order to have large $NN$ scattering lengths. The informative utilities of fine tuning are demonstrated in several examples and naturally interpreted in the underlying theory perspective. In addition, some of the approaches adopted in the recent literature are also addressed in the light of EFT scenario.
\end{abstract}

\end{frontmatter}



\section{Introduction}\label{intro}In the past two decades, the $NN$ systems and nuclear forces have been intensively studied using effective field theory method since Weinberg's proposal\cite{weinberg90,weinberg90b} and the pioneering works in Refs.\cite{ORvKl,ORvK}, which provides nuclear physics with field theoretical foundations in terms of symmetries that characterize low-energy QCD. For more about the exciting achievements and progresses in this area, we refer to the review articles, e.g., \cite{review1a,review1b,review2a,review2b,review3,review4,review5}. However, it is theoretically fair to say that a few stumbling blocks are still in the way towards the complete establishment of the field theoretical foundations for nuclear physics. The renormalization of $NN$ scattering is one of the intriguing issues and has attracted many authors' attention, a comprehensive account of the related literature could be found in the review articles\cite{review1a,review1b,review2a,review2b,review3,review4,review5}. Earlier impetus to this issue was given by the discussions of the problems of Weinberg's power counting in Refs.\cite{KSW1,KSW2,vK}. Since then, a number of approaches have been put forward, ameliorated and discussed, accompanied with some controversies, see, e.g., Refs.\cite{BCP,PBC,Hansen1,Hansen2,Gege1,Gege2,Rho,Steele1,Steele2,EGM1,EGM2,EGM3,Mehen1,Mehen2,Birse1,Birse2,FTT,Chen,FMS1,FMS2,EM1,EM2,EM3,Kuo1,Kuo2,Kuo3,BBSvK,Soto,PVRA1,PVRA2,PVRA3,PVRA4,JFY03,C71,Higa,NTvK,EpelM,YEP1,YEP2,YEP3,I_0,soto-dibaryon1,soto-dibaryon2,LvK,BKV,EpGe,2766,0901.3731,epl2011,Harada,ptPVRA1,ptPVRA2,LY1,LY2,LY3,EGrl1}.
As summarized in Ref.\cite{review3}, in short of complete agreements, there are now two main choices in renormalizing the $NN$ sector: (1) One insists on standard subtraction algorithm of infinities through expanding around some leading component of the $NN$ potential that is first treated nonperturbatively; (2) The other insists on the nonperturbative treatment of the full potential up to the order of truncation, with the renormalization implemented following the method advocated by Lepage\cite{Lepage}.

As noted by many authors\cite{review2a,review2b,review3,JFY03}, the problem is originated from the nonperturbative nature of the low-energy $NN$ scattering: The conventional algorithm for subtraction is only established within perturbation contexts, not guaranteed to work beyond perturbative contexts at all. Of course, the ultimate goals and basic principles of renormalization, should not vary, but the implementation could be context dependent. It is reasonable to anticipate that not all the patterns, contents and scenarios of renormalization could be fully foreseeable in the standard perturbative algorithms. Therefore, the roles played by EFT power counting rules, the implementation of subtractions and the associated wisdoms should all be reexamined and adapted in nonperturbative regimes. The present status about the $NN$ sector in EFT approach implies that we are still in short of a satisfactory framework in nonperturbative contexts. In this regard, the literature could all be understood as various efforts towards the full realization of renormalization in nonperturbative contexts or regimes, and Lepage's proposal could be interpreted as the first conceptual shift in this direction.

Some of the above issues have been touched upon in our previous works\cite{JFY03,C71,I_0,2766,0901.3731,epl2011}. In this report, we wish to present a study of the $NN$ scattering in the coupled channels $^3S_1$$-$$^3D_1$ with a more coherent account of our studies of EFT($\not\!\!\pi$), the extension of the results to higher orders and other channels is a straightforward matter. The renormalization of EFT($\not\!\!\pi$) has been settled using 'perturbative' expansion in Refs.\cite{review1a,review1b,Chen}. Here it is revisited for the availability of closed-form $T$ matrices for exploring crucial notions and contents intrinsic of nonperturbative renormalization. We wish to highlight the nonperturbative components that must be incorporated in any reasonable treatment of EFT for nuclear forces. Our formulation and analysis could also be applied in many other non-relativistic systems that are dominated by short range interactions. We should also note in advance that an EFT upper scale is physical as it actually defines the physical range of an EFT. For example, in EFT($\not\!\!\pi$), the upper scale should be set by pion mass $\Lambda_{(\not\pi)}\sim m_\pi$. In pionfull theory, the upper scale is naturally set by, say, mass of $\rho$ meson, $\Lambda_{(\pi)}\sim m_{\rho}$. The same arguments apply to other low-energy effective field theories.

This report is organized as follows: Section~\ref{sec:1} is devoted to the setups and the rigorous solutions to Lippmann-Schwinger equations (LSE) for $^3S_1$$-$$^3D_1$, with the parametrization of the integrals involved being addressed. In Section~\ref{sec:2}, closed-form $T$ matrices are explored to show why perturbative renormalization cease to apply and what must be done in a nonperturbative implementation of renormalization. Also presented are the scenario notion of EFT and examples of nonperturbative
running couplings. Section~\ref{sec:3} will be devoted to the phenomenological aspects of the closed-form $T$ matrices in various EFT scenarios. Some important scenarios are analyzed with the help of effective range expansion, with the rise and utilities of fine-tunings being explored and interpreted. Preliminary EFT($\not\!\!\pi$) predictions of phase shifts in $^3S_1$ channel are also presented and compared across scenarios. Some discussions and a summary of our report will be given in Section~\ref{sec:4}.
\section{Rigorous solutions in EFT($\not\!\pi$)}\label{sec:1}
\subsection{Preliminary Setups}\label{sec:1.1}Let us start with a standard parametrization for the on-shell partial wave $S$ and $T$ matrices for the triplet $NN$ scattering states with total angular momentum $j$:\bea\label{SYM-param}{\bf S}\equiv\left(\begin{array}{cc}[\cos2\epsilon_j(p)]e^{[2i\delta^{1j}_{j\pm1}(p)]}&i[\sin2 \epsilon_j(p)]e^{i[\delta^{1j}_{j-1}(p)+\delta^{1j}_{j+1}(p)]}\\i[\sin2\epsilon_j(p)]e^{i[\delta^{1j}_{j-1}(p)+\delta^{1j}_{j+1}(p)]}&[\cos2\epsilon_j(p)]e^{[2i\delta ^{1j}_{j\pm1}(p)]}\\\end{array}\right)={\bf I}-\frac{iMp}{2\pi}{\bf T},\eea where $\delta^{1j}_{j\pm1}$ and $\epsilon_j$ denote the phase shifts and mixing angle that depend on the on-shell momentum $p$ of nucleon, $M$ being the nucleon mass. According to Weinberg, the $T$ matrices are obtained through solving the Lippmann-Schwinger equations with the $NN$ potential being systematically constructed using $\chi$PT\cite{weinberg90} through counting the powers of $(p,m_{\pi})$ against the upper scale for the EFT, $\Lambda(\sim0.5$ GeV):\bea\label{LSE}{\bf T}(q^{\prime},q;E)={\bf V}(q^{\prime},q)+\int_k{\bf V}(q^{\prime},k) G_0(k;E){\bf T}(k,q;E),\quad G_0(k;E)\equiv \frac{1}{E-k^2/M+i\epsilon}.\eea

In EFT($\not\!\!\pi$), the $NN$ potentials are contact ones and ${\bf V}$ and ${\bf T}$ in $^3S_1$$-$$^3D_1$ channels could be recast into the following factorized form using the trick of Ref.\cite{PBC}:\bea&&{\bf{V}}(q,q^{\prime})=\left(\begin{array}{cc}V_{ss}&V_{sd}\\V_{ds}&V_{dd}\\\end{array}\right)=\left(\begin{array}{cc}U^T(q^2) \lambda_{ss}U({q^{\prime}}^2)&U^T(q^2)\lambda_{sd}U({q^{\prime}}^2)\\U^T(q^2)\lambda_{ds}U({q^{\prime}}^2)&U^T(q^2)\lambda_{dd}U({q^{\prime}}^2)\\\end{array}\right),\\&& {\bf{T}}(q,q^{\prime};E)=\left(\begin{array}{cc}T_{ss}&T_{sd}\\T_{ds}&T_{dd}\\\end{array}\right)=\left(\begin{array}{cc}U^T(q^2)\tau_{ss}(E)U({q^{\prime}}^2)&U^T(q^2) \tau_{sd}(E)U({q^{\prime}}^2)\\U^T(q^2)\tau_{ds}(E)U({q^{\prime}}^2)&U^T(q^2)\tau_{dd}(E) U({q^{\prime}}^2)\\\end{array}\right),\eea with $U^T(q^2)\equiv(1,q^2,q^4, \cdots)$ being a row vector in terms of external momentum $q$ and $\lambda_{\cdots}$ a matrix of contact couplings. (The energy dependence in the potentials can be removed using unitary transformations\cite{EGM1}.) For example, at truncation order $\mathcal{O}\left(Q^4\right)$ or $\Delta=4$\footnote{Note that this corresponds to the N$^3$LO in pionfull theory.}, we have $U^T(q^2)=(1,q^2,q^4)$, and, $$\lambda_{ss}\equiv\left(\begin{array}{ccc}C_{0;ss}&C_{2;ss}&C_{4;ss}\\C_{2;ss}&\tilde{C}_{4;ss}&0\\ C_{4;ss}&0&0\\\end{array}\right),\ \lambda_{sd}\equiv\left(\begin{array}{ccc}0&0&0\\C_{2;sd}&\tilde{C}_{4;sd}&0\\C_{4;sd}&0&0\\\end{array}\right)=\lambda_{ds}^T,\ \lambda _{dd}\equiv\left(\begin{array}{ccc}0&0&0\\0&C_{4;dd}&0\\0&0&0\\\end{array}\right).$$ The couplings $[C_{n;\ldots}]$ scale like $[C_{n;\ldots}/C_0\sim\Lambda_{(\not\pi)} ^{-n}]$ in naive power counting scheme, with $\Lambda_{(\not\pi)}$ being the upper scale for EFT($\not\!\!\pi$). Note that certain elements of the matrices $[\lambda_{\cdots}]$ vanish at a given order due to truncation, a consequential fact to be explored in Sect.~\ref{sec:2.1}.

Stripping off the $U$ vectors, the Eqs.(\ref{LSE}) can be reduced to four coupled algebraic equations\cite{PBC,C71},\bea\label{algebrTss}\tau_{ss}=\lambda_{ss}+\lambda_ {ss}{\mathcal{I}}(E)\tau_{ss}+\lambda_{sd}{\mathcal{I}}(E)\tau_{ds},\ \tau_{sd}=\lambda_{sd}+\lambda_{sd}{\mathcal{I}}(E)\tau_{dd}+\lambda_{ss}{\mathcal{I}}(E)\tau_{sd}, \ \cdots,\eea with\bea\label{IEdef} {\mathcal{I}}(E)\equiv\int\frac{d^3k}{(2\pi)^3}\frac{U(k^2)U^T(k^2)}{E-k^2/M+i\epsilon},\eea where the energy dependence of the $\tau$'s are self evident and henceforth omitted. The matrix ${\mathcal{I}}(E)$ is furnished with the integrals arising from the convolution with $G_0$, i.e., all the divergences are clearly factorized into this matrix. This fact will yield great convenience for us in the following deductions. The solutions to Eq.(\ref{algebrTss}) are straightforward to find,\bea&&\label{tau-solution1}\tau_{ss}=(1-\tilde{\lambda}_{ss}{\mathcal{I}}(E))^{-1}\tilde{\lambda}_{ss},\ \tau_{sd}=(1-\tilde{\lambda}_{ss} {\mathcal{I}}(E))^{-1}\lambda_{sd}(1-{\mathcal{I}}(E)\lambda_{dd})^{-1},\ \cdots,\\\label{tildelambda1}&&\tilde{\lambda}_{ss}\equiv\lambda_{ss}+\lambda_{sd}{\mathcal{I}} (E)(1-\lambda_{dd}{\mathcal{I}}(E))^{-1}\lambda_{ds},\ \tilde{\lambda}_{dd}\equiv\lambda_{dd}+\lambda_{ds}{\mathcal{I}}(E)(1-\lambda_{ss}{\mathcal{I}}(E))^{-1}\lambda_ {sd}.\eea Each $T_{\cdots}$ matrix could now be obtained from $U^T\tau_{\cdots}U$.

The above results could also be cast into succinct form using the following block matrix notations for $\tau_{xy},\lambda_{xy}$:\bea\underline{\lambda}\equiv\left( \begin{array}{cc}\lambda_{ss}&\lambda_{sd}\\\lambda_{ds}&\lambda_{dd}\\\end{array}\right),\quad\underline{\tau}\equiv\left(\begin{array}{cc}\tau_{ss}&\tau_{sd}\\\tau_ {ds}&\tau_{dd}\\\end{array}\right),\quad\underline{{\mathcal{I}}}(E)\equiv\left(\begin{array}{cc}{\mathcal{I}}(E)&{\bf0}\\{\bf0}&{\mathcal{I}}(E)\\\end{array}\right), \eea then the algebraic LSE's and their solutions read,\bea\label{algsuperLSE}&&\underline{\tau}=\underline{\lambda}+\underline{\lambda}\underline{{\mathcal{I}}}(E) \underline{\tau},\\\label{algsupersolution}&&\underline{\tau}=\left(1-\underline{\lambda} \underline{{\mathcal{I}}}(E)\right)^{-1}\underline{\lambda}.\quad\eea We have verified that Eq.(\ref{algsupersolution}) do reproduce the solutions given in Eqs.(\ref{tau-solution1}) using the formulae given in Appendix A.
\subsection{Parametrization of the matrix ${\mathcal{I}}(E)$}
\label{sec:1.2}The renormalization of the $T$ matrices for $NN$ scattering within EFT$(\not\!\!\pi)$ now boils down to the renormalization of the matrix ${\mathcal{I}} (E)$ to be realized or implemented within nonperturbative context. A generic element of ${\mathcal{I}}(E)$ reads\bea\label{div-n}{\mathcal{I}}_{n}\equiv\int\frac{d^3k} {(2\pi)^3}\frac{k^{2n}}{E-k^2/M}.\eea Such an integral can be parametrized as follows:\bea\label{Jgeneral}&&{\mathcal{I}}_{n}=-\mathcal{I}_0p^{2n}+\sum_{l=1}^nJ_{2l+1} p^{2(n-l)},\quad\mathcal{I}_0\equiv J_0+i\frac{M}{4\pi}p,\quad p\equiv\sqrt{ME},\eea with $J_{\cdots}$ being prescription-dependent parameters (usually constants) at this stage. For example, in the hard cutoff ($\Lambda$) scheme, we have\bea\label{cutoff}J_{0}=\frac{M}{2\pi^2}\left(\Lambda-\frac{p}{2}\ln\frac{\Lambda+p}{\Lambda-p}\right), \quad J_{2l+1}=-\frac{M}{2\pi^2}\frac{\Lambda^{2l+1}}{2l+1}.\eea In dimensional regularization, such integral reads,\bea\label{DR}J_{0}=0,\quad J_{2l+1}=0.\eea In PDS\cite{KSW1,KSW2,Mehen1,Mehen2}, we have\bea\label{PDS}J_{0}=\frac{M}{4\pi}\mu,\quad J_{2l+1}=0.\eea

Actually,we could compute the integral in Eq.(\ref{div-n}) using a simple strategy\cite{D65a,D65b,D65c}: First we differentiate the integral with respect to $E$ or $p^2$ for sufficient times to arrive at a convergent one and carry out the integration,\bea\left(\partial_{p^2}\right)^{n+1}{\mathcal{I}}_{n}=-i\frac{\Gamma\left(n+\frac{3}{2} \right)M}{4\pi\sqrt{\pi}}p^{-1}.\eea Then, upon integrating back indefinitely, the result of Eq.(\ref{Jgeneral}) is exactly reproduced with $J_0$ and $[J_{2m+1},m>0]$ being the corresponding integration constants, which could be seen as a general parametrization of the decoupling effects of underlying
structures\cite{D65a,D65b,D65c,C71}.

Now, the matrix ${\mathcal{I}}(E)$ could be recast into the following succinct form\bea\label{I-short}&&{\mathcal{I}}(E)=-\mathcal{I}_0U(p^2)U^T(p^2)+\sum_{l=1}^\Delta J_{2l+1}\Delta U_l,\\&&\Delta U_1\equiv p^{-2}\int^{p^2}_0dt\frac{d[U(t)U^T(t)]}{dt},\ \Delta U_{l+1}\equiv p^{-2}\int^{p^2}_0dt\frac{d[\Delta U_l(t)]}{dt},\ l\geq1.\eea The concrete expressions for $\Delta U_l$ at order $\Delta=4$ are listed in Appendix B.
\subsection{Closed-form $T$ matrices}\label{sec:1.3}As mentioned above, the closed-form $T$ matrices could be readily obtained by sandwiching the $\tau$'s in Eq.(\ref{tau-solution1}) between the row and column vectors $U^T(q^2)$ and $U({q^\prime}^2)$. After some algebra, we could find following the closed-form on-shell $T$ matrices at order $\Delta=4$:\bea\label{invTss4}&&\frac{1}{T_{ss}(p)}=\mathcal{I}_0+\frac{{\mathcal{N}}_0+\mathcal{I}_0{\mathcal{N}}_1p^4}{{\mathcal{D}}_0+\mathcal{I}_0 {\mathcal{D}}_1 p^4},\quad\frac{1}{T_{dd}(p)}=\mathcal{I}_0+\frac{{\mathcal{N}}_0+\mathcal{I}_0{\mathcal{D}}_0}{\left({\mathcal{N}}_1+\mathcal{I}_0{\mathcal{D}}_1\right) p^4},\\\label{invTsd4}&&\frac{1}{T_{sd}(p)}=\frac{{\mathcal{N}}_0+\mathcal{I}_0\left({\mathcal{D}}_0+{\mathcal{N}}_1p^4\right)+\mathcal{I}_0^2{\mathcal{D}}_1p^4} {{\mathcal{D}}_{sd}p^2}=\frac{1}{T_{ds}(p)},\\\label{keydsd}&&{\mathcal{N}}_1{\mathcal{D}}_0={\mathcal{D}}_{sd}^2+{\mathcal{D}}_1{\mathcal{N}}_0.\eea We note that all the parameters $[{\mathcal{N}}_{\cdots},{\mathcal{D}}_{\cdots}]$ are real polynomials in terms of couplings $[C_{\cdots}]$, $[J_{2m+1},m>0]$ and on-shell momentum $p$, which are all independent of the complex parameter $\mathcal{I}_0$, for detailed expressions, see Appendix C.

As a matter of fact, the functional forms of the $T$ matrices in terms of $[{\mathcal{N}}_{\cdots},{\mathcal{D}}_{\cdots}]$ and $\mathcal{I}_0$ given above hold at any truncation order. To see this, let us invert the matrix $\bf T$ with on-shell entries,\bea\label{Tinv4}{\bf T}^{-1}=\left(\begin{array}{cc}{\mathcal{I}_0}+{\mathcal{N}}_ 1/{\mathcal{D}}_1,&-{\mathcal{D}}_{sd}/({\mathcal{D}}_1p^2)\\-{\mathcal{D}}_{sd}/({\mathcal{D}}_1p^2),&{\mathcal{I}_0}+{\mathcal{D}}_0/({\mathcal{D}}_1p^4)\\\end{array} \right).\eea Then, it is immediate to see that the on-shell unitarity is fulfilled in any prescription at $\Delta=4$, \bea{\bf{T}}^{-1}-({\bf{T}}^{\dag})^{-1}=\frac{iMp} {2\pi}{\bf I},\eea with ${\bf I}$ denoting the $2\times2$ unit matrix. Actually, it is straightforward to prove that this on-shell unitarity rigorously holds at any given order of truncation, for completeness of our presentation, the proof is given in Appendix D. Using this unitarity relation as a starting point, we could also establish that the inverse form of the on-shell super matrix $\bf T$ must take the following form {\em at any given order of potential truncation},\bea\label{TINV}{\bf T}^{-1}= {\mathcal{I}_0}{\bf I}+\left(\begin{array}{cc}\displaystyle{\tilde{\mathcal{N}}}_{ss}/{\tilde{\mathcal{D}}}_{ss},&-{\tilde{\mathcal{N}}}_{sd}/{\tilde{\mathcal{D}}}_{sd} \\-{\tilde{\mathcal{N}}}_{sd}/{\tilde{\mathcal{D}}}_{sd},&{\tilde{\mathcal{N}}}_{dd}/{\tilde{\mathcal{D}}}_{dd}\\\end{array}\right),\eea and $[{\tilde{\mathcal{N}}}_ {\cdots},{\tilde{\mathcal{D}}}_{\cdots}]$ must be ${\mathcal{I}_0}$-independent polynomials in terms of $[C_{\dots}],[J_{n},n>0]$ and $p^2$, again, see Appendix D for a simple proof of this corollary. Another interesting corollary is that the functional dependence of the $T_{ss},T_{dd}$ upon ${\mathcal{I}_0}$ and $[{\mathcal{N}}_ {\cdots},{\mathcal{D}}_{\cdots}]$ as exhibited in Eqs.(\ref{invTss4},\ref{invTsd4}) holds at any truncation order. The proof is also straightforward and will not be presented here, we only mention that Eq.(\ref{keydsd}) is an immediate byproduct of such a calculation.

We note in passing that a small part of the foregoing contents have been sketched in our previous works\cite{I_0,2766,0901.3731,epl2011}. Equipped with these closed-form $T$ matrices, we could better explore the issues around the renormalization of the nonperturbative $T$ matrices and physical issues in the following sections.
\section{Closed form and renormalization}\label{sec:2}
\subsection{Intrinsic mismatch and nonperturbative 'finiteness'}
\label{sec:2.1}Conventionally, UV divergences appear in the local part of a vertex diagram, so local counterterms from couplings can be constructed to remove such divergences. Here in EFT$(\not\!\!\pi)$, this means the divergences in the matrix $\mathcal{I}(E)$ must pair up or 'match' with the contact couplings. Unfortunately, such 'matching' is at least partially lost in the closed-form $T$ matrices. For example, in Eq.(\ref{Tinv4}), the complex parameter $\mathcal{I}_0$ is 'isolated' from ('unmatched' with) any coupling in $\bf{T}^{-1}$. In general, the closed-form $T$ matrices constrain $\mathcal{I}_0$ or $J_0(=\Re({\mathcal{I}_0}))$ to be physical or RG invariant, i.e., $J_0$ could only depend on physical scales, say, the upper scale $\Lambda_{(\not\pi)}$ and/or typical scale $Q$ of EFT($\not\!\pi$)\cite{C71,I_0,2766,0901.3731}:\bea\label{J_0phys}J_0=\frac{M}{4\pi}f_{0}(\Lambda_{(\not\pi)},Q),\eea no longer an ordinary running parameter!

Actually, more parameters in $\mathcal{I}(E)$ become 'unmatched' in the closed-form $T$ matrices beyond leading order. A more transparent way to see this is to invert the algebraic LSE in an uncoupled channel\cite{0901.3731}. It suffices to demonstrate it with the $^1S_0$ channel:\bea\tau^{-1}=\lambda^{-1}-{\mathcal{I}}(E).\eea As no element of ${\mathcal{I}}(E)$ is zero while all the elements in the upper left triangle block of $\lambda^{-1}$ vanish due to truncation, there is an intrinsic mismatch between ${\mathcal{I}}(E)$ and $\lambda^{-1}$, i.e., a mismatch between the ill-definedness in ${\mathcal{I}}(E)$ and the available 'pools' for counterterms from $\lambda^{-1}$. A further mismatch exists between the nonzero entries of $\mathcal{I}(E)$ and $\lambda^{-1}$: the $p$ dependence differs. Letting $\lambda^{-1}$ develop $p$ dependence to match $\mathcal{I}(E)$ would lead to nonlocal time dependence in the local potential $V_{^1S_0}$ and in turn ruin the EFT power counting. Putting more couplings into $\lambda$ to make all elements nonzero would simply break the EFT truncation rules and could not help to remove all the mismatches\cite{0901.3731}. Therefore, conventional counterterms could not succeed due to the tight constraints imposed by the closed-form $T$ matrices, the renormalization has to be implemented otherwise.

One immediate way out is to exploit the virtues of the closed-form $T$ matrices. As $T$ matrices' dependence upon $p$ is physical, the unmatched parameters in $[J_{\cdots}]$ have to be separately determined through physical boundaries or inputs, and this is amusingly guaranteed by the EFT truncation that becomes a virtue at this point: The number of parameters $[J_{\cdots}]$ is actually {\em finite} at any given order of truncation\cite{I_0}, so only finitely many nonperturbative divergences are there to be dealt with, and finitely many boundary conditions or inputs to be imposed at a given order of truncation. It is this nonperturbative 'finiteness' that makes the renormalization of the closed-from $T$ matrices feasible, extending the notion of renormalizability somehow in nonperturbative regimes. In fact, all the divergences involved factorize into 'irreducible' ones that furnish a finite dimensional matrix $\mathcal{I}(E)$, whose rank is controlled by the scaling dimension of the contact potential, not by the number of iterations. As the scaling dimension of a potential with pion exchange is still finite due to truncation, we speculate that this kind of 'finiteness' may also be true in pionfull EFT.

Now the only task left over is to subtract the 'irreducible' divergences and to fix the residual constants using appropriate boundary conditions, which will be addressed in next subsection.
\subsection{Underlying theory perspective and subtraction}
\label{sec:2.2}Let us start with the underlying theory perspective. If a low-energy (LE) process was calculated in a complete theory that underlies an EFT in consideration, then the results must be well defined. In the EFT calculation, the LE projection ($\breve{{\mathcal{P}}}_{\texttt{LE}}$) must be performed before loop integrations in order to arrive at EFT propagators and vertices. The problem is that such LE projection usually does not commute with loop integrations: divergences arise from the  wrong order of operation. At one-loop level, we have\bea\label{CTT}\text{C.T.}\equiv[\breve{{\mathcal{P}}}_{\texttt{LE}},\int d\mu(l)].\eea Rearranging the operations in Eq.(\ref{CTT}), we have,\bea\left\{\breve{{\mathcal{P}}}_{\texttt{LE}}\left[\int d\mu(l)f(l,\cdots)\right]\right\}_{\texttt{UT}}=\left\{\int d\mu(l)\breve {{\mathcal{P}}}_{\texttt{LE}}\left[f(l,\cdots)\right]\right\}_{\texttt{EFT}}+\text{C.T.}\left[f(l,\cdots)\right],\eea with $f(l,\cdots)$ denoting the corresponding integrand. Obviously, the {\em commutator in (\ref{CTT}) just provides the counterterm or subtraction operation} needed in EFT in order to recover the finite integral $\breve{{ \mathcal{P}}}_{\texttt{LE}}\left[\int d\mu(l)f(l,\cdots)\right]$ in UT. So, counterterms or subtractions are 'generated' from the commutator of LE projection and loop integrations that are indispensable in any sensible construction of EFT. Accordingly, subtractions must be implemented at loop level in effect from the underlying theory perspective. In perturbative contexts, subtractions/counterterms could be readily realized through local operators in EFT Lagrangian, while in nonperturbative contexts, such a realization is not generally guaranteed, we have to adhere to subtractions at loop level\cite{Gege1}.

Thus, both the closed-form $T$ matrices and the underlying theory perspective require that subtractions be performed at the level of (loop) integrals in effect through whatever means that makes sense\footnote{In our previous work\cite{C71}, the counterterms in the closed-form $T$ matrices have been termed as 'endogenous' counterterms. Therefore, 'endogenous' counterterms are just the ones that could effectively perform subtractions at loop level in nonperturbative regimes.}. In EFT$(\not\!\!\pi)$, such subtractions are straightforward to perform in the parametrization $\mathcal{I}(E)$, resulting in residual ambiguous parameters (also denoted as $[J_{\cdots}]$) to be fixed through appropriate boundary conditions. This is now a well accepted algorithm in doing the nonperturbative renormalization of EFT\cite{epel1001}. In a sense, the renormalization of Schr\"odinger equations (or Lippmann-Schwinger equations) \'a la Lepage adopted in Refs.\cite{Rho,Steele1,Steele2,EGM1,EGM2,EGM3,EM1,EM2,EM3} can be seen as such instance, with the parameters treated as independent parameters to be fixed separately. Here we note that, perturbative renormalization is done after cut-off independence is achieved, as redefinition is always possible there. In nonperturbative cases, redefinition is usually impossible without ruining the closed form of $T$ matrices, so renormalization is not done with cut-off independence, some parameters must be separately determined.
\subsection{The notion of EFT scenario in nonperturbative regime}
\label{sec:2.3}Basing on the foregoing exploration of the closed-form $T$ matrices, we remark as below: (1) Nonperturbative divergences intrinsically mismatch with the EFT couplings or interactions in the truncated 'space'; (2) Fortunately, the 'unmatched' divergences are finitely many at a given order of potential truncation; (3) The 'unmatched' divergences have to be subtracted at loop level with the residual parameters being physical and independent ones $[J_{\cdots}^{\text{\tiny(phys)}}]$; (4) Then, a novel concept intrinsic of nonperturbative formulation, the 'scenario of EFT' ($\mathcal{S}_{\text{\tiny EFT}}$), naturally emerges in the parametrization of the closed-form $T$ matrices\cite{I_0,epl2011}:\bea\mathcal{S}_{\text{\tiny EFT}}&=&\left[C_{\cdots}(\mu)\right]\oplus\left\{\left[J_{\cdots}^{\text{\tiny(phys)}}\right] \oplus\left[J_{\cdots}(\mu)\right]\right\}\nonumber\\&=&\left\{\left[C_{\cdots}(\mu)\right]\oplus\left[J_{\cdots}(\mu)\right]\right\}\oplus\left[J_{\cdots}^{\text{\tiny
(phys)}}\right]\nonumber\\&=&\left[C^{\text{\tiny(phys)}}_{\cdots}\right]\oplus\left[J_{\cdots}^{\text{\tiny(phys)}}\right],\eea with $\mu$ being the running scale in EFT$(\not\!\!\pi)$. Namely, a scenario consists of EFT couplings and complementary parameters that only arise from loop integrals; (5) EFT power counting could only be manifested through the renormalized EFT couplings, not directly applicable to the bare objects or counterterms in nonperturbative regimes\cite{Gege2}.

Therefore, we anticipate that the scenario notion delineated above must have been incorporated somehow in the EFT descriptions of the $NN$ sector. For example, the separation scale $\lambda$ ($\sim750$ MeV) in Ref.\cite{BKV} essentially plays the roles of the additional parameters complementary to EFT couplings in contrast to the EFT running scale $\mu$ that is of order $m_\pi$, a discrimination that is necessary and natural in the light of EFT scenario. The indispensability of the complementary scenario parameters could in practice also appear in various disguises, see, e.g., Refs.\cite{Rho,Steele1,Steele2,EGM1,EGM2,EGM3,EM1,EM2,EM3,Kuo1,Kuo2,Kuo3,YEP1,YEP2,YEP3,BKV}, or even embodied somehow in different specification of 'dynamical' degrees\cite{soto-dibaryon1,soto-dibaryon2}. Of course, such notion of EFT scenario may not be fully appreciated in the Wilsonian RGE analysis of EFT couplings as it is focused on EFT couplings, not on the full scenario 'space'\cite{Harada}.

So far no deformation or extra construction has been introduced into the EFT framework. All the notions addressed above are natural consequences of the closed-form $T$ matrices unforeseeable in perturbative context. Hence, the original goal of EFT approach--providing field theoretical foundations to nuclear physics--is preserved. The stringent constraints imposed by the closed-form $T$ matrices have not been circumvented, instead, they are directly confronted, exploited and finally turned into virtues within field theoretical framework.
\subsection{Nonperturbative running couplings}\label{sec:2.4}
According to the foregoing notion of EFT scenario: At least some of the EFT couplings develop nonperturbative running behaviors due to intertwining with the running parameters $[J_{\cdots}(\mu)]$. Let us demonstrate it with the coupled channels $^3S_1$$-$$^3D_1$ at order $\Delta=2$, where $J_0$ is the only physical parameter in $[J_{\cdots}]$ that does not run: $\mathcal{N}_1=0,\ \mathcal{N}_0=(1-C_{0;ss}J_3)^2,\ \mathcal{D}_1=-C^2_{2;sd},\ \mathcal{D}_{sd}=C_{2;sd}(1-C_{2;ss}J_3),\ \mathcal {D}_0=\delta_{0;0}+\delta_{0;1}p^2$, with\bea&&\delta_{0;0}=C_{0;ss}+(C^2_{2;ss}+C^2_{2;sd})J_5,\quad\delta_{0;1} =2C_{2;ss}+(C^2_{2;sd}-C^2_{2;ss})J_3.\eea

Then the following combinations are RG invariants besides $J_0$:\bea\alpha_0=\frac{\delta_{0;0}}{\mathcal{N}_0},\quad\alpha_2=\frac{\delta_{0;1}}{\mathcal{N}_0},\quad \beta=\frac{\mathcal{D}_{sd}}{\mathcal{N}_0},\eea as they parametrize physical dependence of the $T$ matrices upon $p$ as below:\bea\frac{1}{T_{ss}}&=&\mathcal{I}_0 +\frac{1}{\alpha_0+\alpha_2p^2-\mathcal{I}_0 \beta^2p^4},\\\frac{1}{T_{dd}}&=&\mathcal{I}_0+\frac{1+(\alpha_0+\alpha_2p^2)\mathcal{I}_0}{-\mathcal{I}_0\beta^2p^4},\\ \frac{1}{T_{sd}}&=&\frac{1+(\alpha_0+\alpha_2p^2)\mathcal{I}_0-\mathcal{I}^2_0\beta^2p^4}{\beta p^2}.\eea Then the running couplings that absorb the running parameters $J_3$ and $J_5$ could be found as below:\bea C_{0;ss}=\left(\alpha_0-\beta^2J_5\right)\xi^{-2}-J_5J^{-2}_3\left(1-\xi^{-1}\right)^2,\quad C_{2,ss}=J^{-1}_3\left(1-\xi^ {-1}\right),\quad C_{2;sd}=\beta\xi^{-1},\eea with $\xi\equiv\sqrt{1+\alpha_2J_3-\beta^2J^2_3}$.

These nonperturbative running couplings possess both IR and UV fixed points in literal sense\bea&&C_{0;ss}^{(\texttt{\tiny IR})}=\alpha_0,\quad C_{2;ss}^{(\texttt{\tiny IR})}=\frac{\alpha_2}{2},\quad C_{2;sd}^{(\texttt{\tiny IR})}=\beta,\quad\quad\\&&C_{0;ss}^{(\texttt{\tiny UV})}=0,\quad C_{2;ss}^{(\texttt{\tiny UV})}=0,\quad C_{2;sd}^ {(\texttt{\tiny UV})}=0.\eea However, these nonperturbative running couplings blow up already at a finite $J_{3}$,\bea J_{3;\pm}=\frac{\alpha_2\pm\sqrt{\alpha^2_2+4\beta ^2}}{2\beta^2}.\eea That means, it does not make sense to let the running scale or the subtraction point go up to UV end, corroborating the fact that this EFT only makes sense below a finite upper scale. In pionfull EFT, it is an extremely challenging task to correctly calculate the contributions from all the intermediate states, especially the sophisticated suppression of higher modes.
\section{Various scenarios of EFT($\not\!\pi$) and phenomenology}\label{sec:3}
\subsection{High- and low-energy behaviors}\label{sec:3.1}First, let us entertain ourselves with some interesting estimates about the high- and low-energy on-shell behaviors of the closed-form $T$ matrices obtained above.

From Appendix C, one could easily read off the following high-energy or UV on-shell behaviors (i.e., $p\rightarrow$ large) at truncation order $\Delta=4$:\bea\frac{1} {T_{ss}}=J_0+i\frac{M}{4\pi}p+o(p^{-2}),\ \frac{1}{T_{dd}}=J_0+i\frac{M}{4\pi}p+o(p^{-2}),\ T_{sd}=T_{ds}=o(p^{-6}).\eea In terms of the parametrization defined in Eq.(\ref{SYM-param}), this is:\bea&&\delta^{(\text{\tiny HE})}_{^3S_1}(p)=\left(n_s+{\textstyle{\frac{1}{2}}}\right)\pi+o(p^{-2}),\quad n_s\in Z,\quad\quad\\&&\delta^ {(\text{\tiny HE})}_{^3D_1}(p)=\left(n_d+{\textstyle{\frac{1}{2}}}\right)\pi+o(p^{-2}),\quad n_d\in Z,\\&&\epsilon^{(\text{\tiny HE})}_1(p)=n_\epsilon\pi+o(p^{-5}),\quad n_\epsilon\in Z.\eea In spite that such behaviors could not be realistic as the EFT($\not\!\!\pi$) description of $NN$ scattering is only valid well below 0.2 GeV, they are still compatible with unitarity. Such behaviors might be reasonable in certain non-relativistic systems that will be studied elsewhere.

Meanwhile, in the low-energy or infrared limit ($p\rightarrow0$), we have\bea\frac{1}{T_{ss}}=J_0+\frac{\nu_{0;0}}{\delta_{0;0}} +o(p),\ \frac{p^4}{T_{dd}}=\frac{\nu_ {0;0}+J_0\delta_{0;0}}{\nu_{1;0}+J_0\delta_{1;0}}+o(p),\ T_{sd}=T_{ds}=o(p^2),\eea and\bea& &\delta_{^3S_1}(0)=n_S\pi,\quad n_S\in Z,\quad\quad\\&&\delta_{^3D_1}(0)=n_D \pi,\quad n_D\in Z,\\&&\epsilon_1(0)=n_E\pi,\quad n_E\in Z.\eea For the realistic $NN$ scattering, we know that $n_S=1,n_D=n_E=0$. Obviously, these behaviors are also compatible with unitarity. More extensive studies of the low-energy behaviors will be presented in terms of effective range expansion (ERE) in Section~\ref{sec:3.4}.

We note that the foregoing behaviors are obtained from the parameters $[{\mathcal{N}}_{\cdots},{\mathcal{D}}_{\cdots}]$ as polynomials of $p^2$ at order $\Delta=4$, where the highest power '$\bar{\omega}$' of $p^2$ can be read off from the expressions listed in Appendix C: $\bar{\omega}_{\mathcal{D}_0}=4,\ \bar{\omega}_{\mathcal{D}_1}= \bar{\omega}_{\mathcal{N}_0}=3,\ \bar{\omega}_{\mathcal{N}_1}=\bar{\omega}_{\mathcal{D}_{sd}}=2$. Then it is rational to expect that the following ranking should hold at higher order of truncations:\bea&&\bar{\omega}_{\mathcal{D}_0}>\bar{\omega}_{\mathcal{D}_1},\ \bar{\omega}_{\mathcal{N}_0}>\bar{\omega}_{\mathcal{N}_1},\ \bar{\omega} _{\mathcal{D}_0}>\bar{\omega}_{\mathcal{N}_0},\quad\min \{\bar{\omega}_{\mathcal{D}_1},\bar{\omega}_{\mathcal{N}_0}\}>\bar{\omega}_{\mathcal{D}_{sd}}.\eea With these rankings, one may convince oneself that the above limiting behaviors should qualitatively persist at higher orders of truncation.
\subsection{Scale hierarchy and scenarios}\label{sec:3.2}In the following, we examine the behaviors of the closed form $T$ matrices obtained above in the light of EFT scenario and extend the preliminary analysis given in Ref.\cite{epl2011}. Here and below, the subscript '$(\not\!\pi)$' in $\Lambda_{(\not\pi)}$ is omitted to avoid heavy symbolism.

Generically, the scenario parameters $[C_{\cdots}]$ and $[J_{\cdots}]$ depend on the ratio $\epsilon\equiv{Q}/{\Lambda}$ or ${\mu}/{\Lambda}$ that stipulates EFT expansion due to scale hierarchy: $\mu\sim Q,\ Q\ll\Lambda$.  For the realistic $NN$ scattering in the realm of EFT($\not\!\!\pi$) where $\Lambda\simeq m_\pi,\ a^{-1} ({^3S_1})\simeq36.4\text{MeV},\ a^{-1} ({^1S_0})\simeq-8.3\text{MeV}$, etc., we could envisage the following scale hierarchy:\bea\label{NN-hierarchy}\epsilon\simeq {\textstyle\frac{1}{4}}:\quad a^{-1}({^3S_1})\simeq{o(\epsilon)\Lambda},\quad a^{-1}({^1S_0})\simeq-{o(\epsilon^2)\Lambda},\quad\cdots.\eea To proceed, we introduce the following dimensionless parameters for the couplings and $J_{\cdots}$:\bea C_{2n;\cdots}=\frac{4\pi}{M}\frac{\tilde{c}_{2n;\cdots}(\epsilon)}{2^{n}\Lambda^{2n+1}}, \quad\label{dimlessJk}J_{2k+1}=\frac {M\mu^{2k+1}}{4\pi}\tilde{j}_{2k+1}(\epsilon),\quad\eea where $\tilde{c}_{2n;\cdots}$ may be multiplied by powers of '2' due to our convention. In complementary parameters $[J^{\text{\tiny(phys)}}_{2k+1}]$, '$\mu$' is replaced with '$Q$'. Below, we will consider the following three typical scenarios for simplicity:\bea\label{SCA}\text{A:}&&\tilde{c}_{2n;\cdots}\sim{\mathcal{O}}(1);\ \tilde{j}_{2k+1}\sim{\mathcal{O}}(1);\ J_0\sim\frac{M}{4\pi}Q;\\\label{SCB}\text{B:} &&\tilde{c}_{2n;\cdots}\sim\frac{{\mathcal{O}}(1)}{\epsilon^{n+1}};\ \tilde{j}_{2k+1}\sim{\mathcal{O}}(1);\ J_0\sim\frac{M}{4\pi}Q;\quad\quad\\\label{SCC}\text{C:}&& \tilde{c}_{2n;\cdots}\sim{\mathcal{O}}(1);\ \tilde{j}_{2k+1}\sim{\mathcal{O}}(1);\ J_0\sim\frac{M}{4\pi}\Lambda.\eea Obviously, scenario A will lead to natural ERE parameters, hence a natural scenario. Scenario B incorporates unconventional power counting of couplings, it will indeed lead to unnatural scattering behaviors. Scenario C is nearly the same as scenario A except $J_0(=\text{Re}[\mathcal{I}_0])$. This is because as a physical or RG invariant parameter (see Eq.(\ref{J_0phys})), $J_0$ could simply be a function of the physical upper scale $\Lambda$ only\cite{C71,I_0,0901.3731}. Actually, scenario C is 'natural' in the sense that all the scales involved are 'naturally' sized, but it could also lead to unnatural scattering lengths for $S$-waves upon reasonable fine-tunings, see below. In a sense, scenario C provides a natural foundation to the EFT treatments (see reviews\cite{review1a,review1b,review2a,review2b,review3,review4,review5}) that employ various forms of fine tuning. In our view, the field theoretical origin of the complexity in nuclear physics just lies in the nonperturbative regime of EFT renormalization, or, in the nonperturbative scenario of EFT. Further remarks on fine tuning will be given in Section~\ref{sec:3.5}.

Here, some remarks are in order: 1) In principle, the running $\mu$ can be any thing below the upper scale $\Lambda$ in an EFT, so, $\epsilon(=\mu/\Lambda)\in(0,1)$; 2) In phenomenologies, $\mu$ is usually sized as the typical momentum $Q$, i.e., $\epsilon\sim\frac{1}{4}$ as in Eq.(\ref{NN-hierarchy}); 3) Peculiar choices with $\mu\sim\Lambda$ are theoretically possible, resulting in more sophisticated scenarios which will be studied elsewhere in future.
\subsection{$T$ matrices in various scenarios}\label{sec:3.3}
Obviously, the patterns of fine tuning would differ across scenarios. In the following, the fine tuning in a scenario will be defined in terms of $[\tilde{c}_{\cdots} (\epsilon)]$ as below:\bea\frac{\tilde{c}_{2n;\cdots}(\epsilon)}{|\tilde{c}_{2n;\cdots}(0)|}=\pm1+o(\epsilon^{\sigma}),\quad\sigma\in(0,\kappa),\eea where $\kappa$ denotes the smallest exponent in $\epsilon$ expansion of the contributions one order higher than the coupling $C_{2n;\cdots}$ to the coefficients $[\delta_{\cdots}]$ (C.f. Appendix C). Evidently, $\kappa$ varies with scenario and larger $\kappa$ means larger capacity for fine tuning, less sensitivity to higher order contributions, and finally more credits for EFT approach in the corresponding scenario. It could be obtained through studying the $\epsilon$ dependence of the $T$ matrices, which will be demonstrated below with a simple choice $\tilde{j}_{2k+1}=1$ and $\mu=Q=o(\epsilon)\Lambda$.
\subsubsection{Scenario A}\label{sec:3.3.1}In this scenario, we have,\bea\frac{4\pi}{M\Lambda}T^{-1}_{ss}=\epsilon+\frac{ip}\Lambda+\frac{1+o(\epsilon^3)+\frac{\epsilon ^3p^2}{\Lambda^2}{\mathcal{O}}\left(1+o(\epsilon^3)\right)+\cdots}{\tilde{c}_{0;ss}+o(\epsilon^5)+\frac{p^2}{\Lambda^2}\mathcal{O}\left(1+o(\epsilon^3)\right)+\cdots}, \quad\cdots.\eea Closer study shows that for $\tilde{c}_{0;ss}$ in this scenario: $\kappa_{\text{\tiny A}}=3$, a large capacity that renders fine tuning stable against higher order corrections. However, this favorable capacity is useless in this natural scenario at all, as can be seen from the ERE parameters in $^3S_1$ channel:\bea {a^{-1}}\sim{-\tilde{c}^{-1}_{0;ss}}\Lambda\sim\mathcal{O}(1)\Lambda,\quad r_e\sim2\tilde{c}_{2;ss}\Lambda^{-1}\sim\mathcal{O}(1)\Lambda^{-1}.\eea That is, the ERE parameters are insensitive to fine tuning in scenario A.
\subsubsection{Scenario B}\label{sec:3.3.2}In this scenario that mimics KSW\cite{KSW1} scheme of couplings within the realm of EFT$(\not\!\!\pi)$, we have,\bea\frac{4\pi} {M\Lambda}T^{-1}_{ss}=\epsilon+\frac{ip}{\Lambda}+\frac{1+o(\epsilon)+\frac{p^2}{\Lambda^2}{\mathcal{O}}(1+o(\epsilon))+\cdots}{\tilde{c}_{0;ss}+o(\epsilon)+\frac{p^2} {\epsilon\Lambda^2}\mathcal{O}(1+o(\epsilon))+\cdots},\quad\cdots.\eea Here, we find that, $\kappa_{\text{\tiny B}}=1$ at least for $\tilde{c}_{0;ss}$, less favorable capacity for fine tuning for this scenario. Nevertheless, we could still achieve unnatural scattering length in $^3S_1$ channel due to unconventional couplings:\bea&& a^{-1}\sim-\left(\epsilon+\tilde{c}^{-1}_{0;ss}\right)\Lambda\sim o(\epsilon)\Lambda,\quad r_e\sim2\epsilon^2\tilde{c}_{2;ss}\Lambda^{-1}\sim\mathcal{O}(1)\Lambda^{-1}. \eea However, this scenario would lead to other unnaturally large ERE parameters in $S$-channels also due to the unconventional rating of couplings, see next subsection.
\subsubsection{Scenario C}\label{sec:3.3.3}In this scenario, we have,\bea\frac{4\pi}{M\Lambda}T^{-1}_{ss}=1+\frac{ip}{\Lambda}+\frac{1+o(\epsilon^3)+\frac{\epsilon^3p^2} {\Lambda^2}{\mathcal{O}}\left(1+o(\epsilon^3)\right)+\cdots}{\tilde{c}_{0;ss}+o(\epsilon^5)+\frac{p^2}{\Lambda^2}\mathcal{O}\left(1+o(\epsilon^3)\right)+\cdots},\quad \cdots.\eea Here, $\kappa_{\text{\tiny C}}=3$ for $\tilde{c}_{0;ss}$, a large capacity for fine tuning that is truly pivotal for producing large scattering length in this scenario. That is, we could have,\bea&&a^{-1}\sim-\left(1+\tilde{c}^{-1}_{0;ss}\right)\Lambda\sim o(\epsilon^{\sigma})\Lambda,\quad r_e\sim2\tilde{c}_{2;ss}\Lambda^{-1} \sim\mathcal{O}(1)\Lambda^{-1},\eea with the fine tuning $\tilde{c}_{0;ss}=-1-o(\epsilon^\sigma),\ \sigma\in(0,3)$. For the realistic $^3S_1$ scattering, it suffices to choose $\sigma=1$ so that $a^{-1}\sim o(\epsilon)\Lambda$. Thus, in the realm of EFT$(\not\!\pi)$, a natural effective range $r_e$ and an unnatural scattering length in $^3S_1$ channel could be 'naturally' achieved in scenario C. To see more rationalities, it is instructive to compute and compare more ERE parameters across various scenarios. This will be done below.
\subsection{Effective range expansion in various scenarios and PSA data}\label{sec:3.4} The standard ERE in $L$-wave is defined as below:\bea p^{2L+1}\cot\delta_{L}(p) =-a^{-1}+{\textstyle\frac{1}{2}}r_ep^2+{\textstyle\sum_{n=2}^\infty}v_np^{2n}.\eea In $^3S_1$$-$$^3D_1$, one could arrive at the following low energy relations using Eq.(\ref{SYM-param}):\bea p\ \cot\delta_s(p)&=&-\frac{4\pi}{M}\left\{\Re\left[T_{ss}^{-1}\right]+o\left(p^{10}\right)\right\},\\p^5\cot\delta_d(p)&=&-\frac{4\pi}{M}\left \{p^4/\Re\left[T_{dd}\right]+o\left(p^6\right)\right\},\eea that means, we could compute the ERE parameters up to $v_4$ in $^3S_1$ with $\Re[T^{-1}_{ss}]$, and up to $v_2$ in $^3D_1$ with $\Re[T_{dd}]$. The results will be rational functions in terms of $[\nu_{\cdots},\delta_{\cdots}]$ and $J_0$, which could be further expanded in terms of $\epsilon$.
\begin{table}[h]
\caption{Naturalness(N)/unnaturalness(U) of ERE parameters in $^3S_1$$-$$^3D_1$: Tuning I}\label{tab:1}
\begin{center}\begin{tabular}{|c|c|c|c|}\hline\hline ERE&Scenario A &Scenario B&Scenario C\\\hline$S$: $\Lambda\cdot a$&${\mathcal{O}}(1+o(\epsilon))$&$\epsilon^{-2} \mathcal{O}(1+o(\epsilon))$&$\epsilon^{-1}\mathcal{O}(1+o(\epsilon))$\\$\Lambda\cdot r_e$&$2\tilde{c}_{2;ss}+o(\epsilon)$&$2\epsilon^2\tilde{c}_{2;ss}+o(\epsilon)$&$2 \tilde{c}_{2;ss}+o(\epsilon)$\\$\Lambda^3\cdot v_2$&${\mathcal{O}}(1+o(\epsilon))$&$\epsilon^{-1}{\mathcal{O}}(1+o(\epsilon))$&${\mathcal{O}}(1+o(\epsilon))$\\$\Lambda^5 \cdot v_3$&${\mathcal{O}}(1+o(\epsilon))$&$\epsilon^{-2}{\mathcal{O}}(1+o(\epsilon))$&${\mathcal{O}}(1+o(\epsilon))$\\$\Lambda^7\cdot v_4$&${\mathcal{O}}(1+o(\epsilon)) $&$\epsilon^{-3}{\mathcal{O}}(1+o(\epsilon))$&${\mathcal{O}}(1+o(\epsilon))$\\\hline$D$: $\Lambda^5\cdot a$&${\mathcal{O}}(1+o(\epsilon))$&$\epsilon^{-4}\mathcal{O} (1+o(\epsilon))$&$\epsilon^{-1}\mathcal{O}(1+o(\epsilon))$\\$\Lambda^{-3}\cdot r_e$&$2{\mathcal{O}}(1+o(\epsilon))$&$2{\mathcal{O}}(1+o(\epsilon))$&$2\epsilon^{-1} {\mathcal{O}}(1+o(\epsilon))$\\$\Lambda^{-1}\cdot v_2$&${\mathcal{O}}(1+o(\epsilon))$&$\epsilon^{-3}{\mathcal{O}}(1+o(\epsilon))$&$\epsilon^{-2}{\mathcal{O}} (1+o(\epsilon))$\\\hline\hline\end{tabular} \end{center}\end{table}
\begin{table}[h]
\caption{Naturalness(N)/unnaturalness(U) of ERE parameters in $^3S_1$$-$$^3D_1$: Tuning II}\label{tab:2}
\begin{center}\begin{tabular}{|c|c|c|}\hline\hline ERE&Scenario A&Scenario C\\\hline$S$: $\Lambda\cdot a$&${\mathcal{O}}(1+o(\epsilon))$&$\epsilon^{-2}\mathcal{O} (1+o(\epsilon))$\\$\Lambda\cdot r_e$&$2\tilde{c}_{2;ss}+o(\epsilon^2)$&$2\tilde{c}_{2;ss}+o(\epsilon^2)$\\$\Lambda^3\cdot v_2$&${\mathcal{O}}(1+o(\epsilon))$&${\mathcal {O}}(1+o(\epsilon^2))$\\$\Lambda^5\cdot v_3$&${\mathcal{O}}(1+o(\epsilon))$&${\mathcal{O}}(1+o(\epsilon^2))$\\$\Lambda^7\cdot v_4$&${\mathcal{O}}(1+o(\epsilon))$&$ {\mathcal{O}}(1+o(\epsilon^2))$\\\hline$D$: $\Lambda^5\cdot a$&${\mathcal{O}}(1+o(\epsilon))$&$\epsilon^{-2}\mathcal{O}(1+o(\epsilon))$\\$\Lambda^{-3}\cdot r_e$&$2 {\mathcal{O}}(1+o(\epsilon))$&$2\epsilon^{-2}{\mathcal{O}}(1+o(\epsilon))$\\$\Lambda^{-1}\cdot v_2$&${\mathcal{O}}(1+o(\epsilon))$&$\epsilon^{-4}{\mathcal{O}} (1+o(\epsilon))$\\\hline\hline\end{tabular}\end{center}\end{table}
\begin{table}[ht]\caption{Naturalness(N)/unnaturalness(U) of ERE parameters in $^1S_0$}\label{tab1s0}
\begin{center}\begin{tabular}{|c|c|c|c|}\hline\hline ERE& Scenario A&Scenario B&Scenario C\\\hline$\Lambda\cdot a$&${\mathcal{O}}(1+o(\epsilon))$&$\epsilon^{-2}\mathcal {O}(1+o(\epsilon))$&$\epsilon^{-2}\mathcal{O}(1+o(\epsilon))$\\$\Lambda\cdot r_e$&$2\tilde{c}_{2}+o(\epsilon)$&$2\epsilon^2\tilde{c}_{2}+o(\epsilon)$&$2\tilde{c}_{2} +o(\epsilon^2)$\\$\Lambda^3\cdot v_2$&${\mathcal{O}}(1+o(\epsilon))$&$\epsilon^{-1}{\mathcal{O}}(1+o(\epsilon))$&${\mathcal{O}}(1+o(\epsilon^2))$\\$\Lambda^5\cdot v_3$& ${\mathcal{O}}(1+o(\epsilon))$&$\epsilon^{-2}{\mathcal{O}}(1+o(\epsilon))$&${\mathcal{O}}(1+o(\epsilon^2))$\\$\Lambda^7\cdot v_4$&${\mathcal{O}}(1+o(\epsilon))$&$ \epsilon^{-3}{\mathcal{O}}(1+o(\epsilon))$&${\mathcal{O}}(1+o(\epsilon^2))$\\\hline\hline\end{tabular}\end{center}\end{table}
\begin{table}[ht]\caption{ERE parameters in $S$-waves: PSA data.}\label{tabPSA}
\begin{center}\begin{tabular}{|c||c|c||c|c|}\hline\hline ERE&$^3S_1$: data&scaling&$^1S_0$: data&scaling\\\hline$\tilde{\Lambda}\cdot a$&$(0.26)^{-1}$&$\varepsilon^{-1} {\mathcal{O}}(1)$&$-(0.06)^{-1}$&$\varepsilon^{-2}{\mathcal{O}}(1)$\\$\tilde{\Lambda}\cdot r_e$&$(0.81)^{-1}$&${\mathcal{O}}(1)$&$(0.53)^{-1}$&$2{\mathcal{O}}(1)$\\ $\tilde{\Lambda}^3\cdot v_2$&$(4.13)^{-3}$&$\varepsilon^3{\mathcal{O}}(1)$&$-(1.81)^{-3}$&$\varepsilon^{\frac{5}{4}}{\mathcal{O}}(1)$\\$\tilde{\Lambda}^5\cdot v_3$&$ (1.53)^{-5}$&$\varepsilon^{\frac{3}{2}}{\mathcal{O}}(1)$&$(1.07)^{-5}$&${\mathcal{O}}(1)$\\$\tilde{\Lambda}^7\cdot v_4$&$-(1.16)^{-7}$&$\varepsilon^{\frac{3}{4}} {\mathcal{O}}(1) $&$-(0.92)^{-7}$&${\mathcal{O}}(1)$\\\hline\hline\end{tabular}\end{center}\end{table}

The detailed $\epsilon$ dependence of the ERE parameters is determined by the fine tuning in the corresponding scenario. The results in $\epsilon$ expansion are summarized in Tables~\ref{tab:1} and~\ref{tab:2}, where the following two primary fine tuning patterns (tuning I and II) for the leading coupling $\tilde{c}_ {0;ss}$ are demonstrated as we are mainly concerned with $S$-wave scattering lengths:\bea&\text{Tuning I:}&\ \tilde{c}_{0;ss}\sim-1-o(\epsilon)\ (\text{scenario A,C}),\quad\epsilon \tilde{c}_{0;ss}\sim-1-o(\epsilon)\ (\text{scenario B});\nonumber\\&\text{Tuning II:}&\ \tilde{c}_{0;ss}\sim-1-o(\epsilon^2)\ (\text{scenario A,C}).\nonumber\eea Note that tuning II is simply forbidden by $\kappa_{\text{\tiny B}}=1$ in scenario B. Fine tuning of higher couplings will be considered in Section~\ref{sec:3.5} for higher ERE parameters. Note that in order to yield a scattering length of order $(\epsilon\Lambda)^{-1}$ in scenario B, one should use $\epsilon\tilde{c}_{0;ss}\sim+1+o(\epsilon)$ instead of tuning I, with the rest being essentially not affected. The $^1S_0$ results are also presented here in Table~\ref{tab1s0}, where to yield a much larger $(\sim\epsilon^{-2})$ scattering length, the tuning $\epsilon\tilde{c}_{0}\sim-1+o(\epsilon)$ is used in scenario B while in scenario C we use $\tilde{c}_{0}\sim-1 +o(\epsilon^2)$. At order $\Delta=4$, the higher ERE parameters ($v_3,v_4$ in $S$ channels, etc) are less trustworthy and listed here only for reference.

From Tables~\ref{tab:1}$-$\ref{tab1s0}, it is obvious that scenario A characterizes systems with natural scattering behaviors, while the rest two account for unnatural systems. We also presented in Table~\ref{tabPSA} the analysis of empirical ERE parameters in $^3S_1$ and $^1S_0$ channels using the PSA data\cite{EGM3}, with the upper scale $\tilde{\Lambda}$ and scaling parameter $\varepsilon$ taken to be $m_{\pi^{\pm}}$ and $\frac{1}{4}$, respectively. Then we see the huge 'gaps' between PSA data and scenario B:\bea{^3S_1}:&&\frac{v_{2;B}}{v_{2;P}}\sim\epsilon^{-4},\ \frac{v_{3;B}}{v_{3;P}}\sim\epsilon^{-\frac{7}{2}},\ \frac{v_{4;B}}{v_{4;P}}\sim\epsilon^{-\frac{15} {4}},\quad\quad\\{^1S_0}:&&\frac{v_{2;B}}{v_{2;P}}\sim\epsilon^{-\frac{9}{4}},\ \frac{v_{3;B}}{v_{3;P}}\sim\epsilon^{-2},\ \frac{v_{4;B}}{v_{4;P}}\sim\epsilon^{-3},\eea with subscript '$_{\cdots;B}$' for scenario B while '$_{\cdots;P}$' for PSA. The 'gaps' are smaller in scenario C:\bea\label{gaps}{^3S_1}:&&\frac{v_{2;C}}{v_{2;P}}\sim \epsilon^{-3},\ \frac{v_{3;C}}{v_{3;P}}\sim\epsilon^{-\frac{3}{2}},\ \frac{v_{4;C}}{v_{4;P}}\sim\epsilon^{-\frac{3}{4}},\quad\quad\\{^1S_0}:&&\frac{v_{2;C}}{v_{2;P}}\sim \epsilon^{-\frac{5}{4}},\ \frac{v_{3;C}}{v_{3;P}}\sim\epsilon^{0},\ \frac{v_{4;C}}{v_{4;P}}\sim\epsilon^{0}.\eea Thus, scenario B seems to be disfavored by the PSA data. In $^1S_0$, the agreement between scenario C and PSA data is almost complete.

The numbers in Tables~\ref{tab:1}$-$\ref{tab1s0} have been derived with primary fine-tunings of the leading couplings $C_{0;\cdots}$ only, not quite informative about higher ERE form factors. Actually, higher ERE form factors $[v_k,k\geq2]$ involve more higher couplings at the 'leading' order of $\epsilon$ expansion, so cancellation amongst the couplings involved may occur, reducing their magnitudes. As will be shown in Section~\ref{sec:3.5}, this could indeed happen in scenario C. In scenario B, however, it is hard to achieve the reduction of magnitudes due to the following tension: (1) On the one hand, the huge 'gaps' would require much larger capacity for fine tuning; (2) On the other hand, the actual capacity is only marginal, $\kappa_{\text{\tiny B}}=1$, which obviously stems from the unconventionally large couplings in scenario B. In this regard, scenario B, or a scenario with unconventional power counting of couplings, is strongly disfavored in the EFT description of $NN$
scattering.

More than a decade ago, treating pion exchanges perturbatively in $NN$ scattering using KSW scheme was shown in Refs.\cite{Hansen1,Hansen2} to lead to large ERE parameters (Table~\ref{CohenLET}), in qualitative agreement with what we found in scenario B. Therefore, the scenarios with unconventionally large couplings do seem to be pathological choices for $NN$ scattering in lower partial waves.
\begin{table}[h]\caption{Low energy theorems from perturbative pions}\label{CohenLET}
\begin{center}\begin{tabular}{|c|c|c|c|c|}\hline\hline ERE&$^3S_1$&scaling&$^1S_0$&scaling\\\hline$v_2(\text{fm}^3)$&$-0.95$&$\varepsilon^{\frac{11}{14}}\frac{{\mathcal {O}}(1)}{\tilde{\Lambda}^{3}}$&$-3.3$&$\varepsilon^{-\frac{1}{9}}\frac{{\mathcal{O}}(1)}{\tilde{\Lambda}^{3}}$\\$v_3(\text{fm}^5)$&$+4.6$&$\varepsilon^{\frac{1}{7}}\frac {{\mathcal{O}}(1)}{\tilde{\Lambda}^{5}}$&$+17.8$&$\varepsilon^{-\frac{5}{6}}\frac{{\mathcal{O}}(1)}{\tilde{\Lambda}^{5}}$\\$v_4(\text{fm}^7)$&$-25$&$\varepsilon^{-\frac {4}{7}}\frac{{\mathcal{O}}(1)}{\tilde{\Lambda}^{7}}$&$-108$&$\varepsilon^{-\frac{13}{8}}\frac{{\mathcal{O}}(1)}{\tilde{\Lambda}^{7}}$\\\hline\hline\end{tabular} \end{center}\end{table}
\subsection{Fine tuning and hidden structures}\label{sec:3.5}
\subsubsection{A small $v_2({^3S_1})$ in scenario C}\label{sec:3.5.1}Let us first show how to achieve a small $v_2$ of $^3S_1$ channel in scenario C with fine tuning. Examining the detailed expression of $v_2$ in scenario C in leading orders of $\epsilon$ expansion:\bea\label{smallv_2}\Lambda^3\cdot v_2=\frac{2\tilde{c}_{4;ss}+\tilde {\tilde{c}}_{4;ss}-4{\tilde{c}}^2_{2;sd}}{4\tilde{c}^2_{0;ss}}-\frac{\tilde{c}^2_{2;ss}}{\tilde{c}^3_{0;ss}}+o(\epsilon^3),\quad\eea it is evident that the higher couplings $\tilde{c}_{2;ss}$, $\tilde{c}_{2;sd}$, ${\tilde{c}}_{4;ss}$ and $\tilde{\tilde{c}}_{4;ss}$ also possess a large capacity $\kappa_{\text{\tiny C}}=3$. Then, a small $v_2$ of order '$o(\epsilon^3)$' would result if the couplings on the right hand side of Eq.(\ref{smallv_2}) cancel out against each other up to order $o(\epsilon^3)$. Considering $v_2$ alone here, this could be achieved in a number of ways. For example, the following choices in combination with tuning I for $\tilde{c}_{0;ss}$ could result in a $v_2$ of the required size:\bea&&\left\{\begin{array}{l}\tilde{c}_{0;ss}\sim-1-o(\epsilon)\\{\tilde{c}}_{2;ss}\sim1-o(\epsilon^2),\ {\tilde{c}}_{2;sd}\sim1-o(\epsilon^2)\\{\tilde{c}}_{4;ss}\sim1+o(\epsilon)-o(\epsilon^2),\ \tilde{\tilde{c}}_{4;ss}\sim2[-1+o(\epsilon)-o(\epsilon^2)]\end{array}\right\} \quad\quad\quad\\&&\Longrightarrow\Lambda^3v_2\sim o(\epsilon^3).\eea

Of course, other ERE parameters may impose additional fine tuning requirements for the couplings, which would further constrain the couplings involved and the choices of fine tuning, i.e., the couplings must be correlated somehow with each other.
\subsubsection{Combined constraints}\label{sec:3.5.2}Below we provide an instance of deriving constraints or correlations for the couplings through combined considerations of the fine tuning requirements from the $^1S_0$ and $^3S_1$ channels in scenario C. To proceed, we adopt the following decomposition of contact $NN$ potential at the leading order of truncation\cite{kaiser,epelthesis}:\bea V_{NN}^{(0)}=C_{c}+(\bm{\tau}_1\cdot\bm{\tau}_2)C_{I;c}+(\bm{\sigma}_1\cdot\bm{\sigma}_2)C_{t} +(\bm{\tau}_1\cdot\bm{\tau}_2)(\bm{\sigma}_1\cdot\bm{\sigma}_2)C_{I;t}.\eea In partial wave representation, we have,\bea C_{^1S_0}=C_{c}+C_{I;c}-3(C_{t}+C_{I;t}),\quad C_{^3S_1}=C_{c}+C_{t}-3(C_{I;c}+C_{I;t}).\eea

From Tables~\ref{tab:1} and~\ref{tab1s0}, it is clear that the following tuning is required in scenario C:\bea\frac{M\Lambda}{4\pi}C_{^1S_0}=-1+o(\epsilon^2),\quad\frac {M\Lambda}{4\pi}C_{^3S_1}=-1-o(\epsilon),\eea which in turn leads to the following correlations among $[C_c$, $C_t$, $C_{I;c}$, $C_{I;t}]$ at the leading order of $\epsilon$ expansion:\bea&&C_t(0)=C_{I;c}(0),\\&&C_c(0)=2C_{I;c}(0)+3C_{I;t}(0)-1.\eea In the EFT approach to $NN$ scattering, these relations follow as corollaries in scenario C. They may be tested with lattice computation. Furthermore, if one could assume that $\frac{M\Lambda}{4\pi}C_c(0)=\frac{M\Lambda}{4\pi}C_t(0)=1$, then the above correlations would imply that $C_{I;t}$ is suppressed by at least one order of $\epsilon$: $C_{I;t}(0)=0,\ C_{I;t}=\frac{4\pi}{M\Lambda}o(\epsilon^a),\ a\geq1.$
\subsubsection{Interpretation}\label{sec:3.5.3}Now we make some attempts at interpreting the foregoing deductions. We have shown that fine tuning in combination with phenomenological requirements can lead us to arrive at quite some 'orders' or 'structures' hidden in the EFT couplings. These hidden 'regularities', should they be correct or trustworthy, must come from certain 'structures' or 'symmetry' contents of the underlying theory.

Let us elaborate. It is natural to expect that the contact couplings of EFT($\not\!\!\!\pi$) should be proportional to $\frac{4\pi}{M}m_\pi^{-n}$, and the constraints must be reflecting the structures of the pion exchange amplitudes in low energy expansion, hence the structures of broken chiral symmetry of chiral perturbation theory or QCD. As a matter of fact, generic arguments may go as below: Each coupling in an EFT is a simplified projected version of an amplitude defined in the underlying theory. Such amplitudes must be constrained by fundamental 'symmetries', hence 'correlated' with each other. Translated into the EFT language, then, we end up with EFT scenario parameters that are correlated with each other. Thus, the correlations among the EFT couplings are nothing else but reflections of the regular 'structures' in the underlying theory. Not knowing the details of the underlying theory, we have to constrain the scenario parameters with empirical data or physical inputs, which is an indispensable step in the EFT approach. In particular, in the scenario C considered here, fine tuning is a 'fine' 'component' of
EFT calculations, naturally driven and soundly supported by empirical data.
\subsection{Phase shifts predictions and scenarios of EFT($\not\!\pi$)}\label{sec:3.6}In this subsection, we present some preliminary predictions of the phase shifts in $^3S_1$ channel using the closed-form $T$ matrices obtained above.  We have chosen to put that, $J_{2n+1}=0,\ \forall n>0,$ to make the numerical work simple. The couplings are determined through fitting to the PSA curve at the low energy end $\texttt{T}_{\texttt{\tiny lab}}\in(0,3]$ MeV. We computed two situations: (a) $\frac{4\pi}{M}J_0=35$ MeV and (b) $\frac{4\pi}{M}J_0=138$ MeV, which simulate scenario B and C respectively. The couplings of $^3S_1$ channel and their scaling behaviors are presented in Table~\ref{tab:phase} ($\mu=1.4$ MeV in scenario B), other couplings and more results will be presented in a separate report in the near future\cite{Phaseshifts}.
\begin{table}[h]\caption{$^3S_1$ couplings fitted with $\texttt{T}_{\texttt{\tiny lab}}\in(0,3]$ MeV.}\label{tab:phase}
\begin{center}\begin{tabular}{|c||cc||cc|}\hline\hline Scenario ($\Lambda=m_\pi$)&B&$(J_0\propto35$)\quad\quad&C&$(J_0\propto138$)\quad\quad\\\hline$C_{0;ss}($MeV$^ {-2}$)&$+9.56\times10^{-3}$&$1.00{\mathcal{O}}\left({\textstyle\frac{4\pi}{M\mu}}\right)$&$-1.32\times10^{-4}$&$1.36\mathcal{O}\left({\textstyle\frac{4\pi}{M\Lambda}} \right)$\\\hline$C_{2;ss}($MeV$^{-4}$)&$+2.21\times10^{-8}$&$0.04{\mathcal{O}}\left({\textstyle\frac{4\pi}{M\mu^2\Lambda}}\right)$&$+2.79\times10^{-9}$&$0.55\mathcal{O} \left({\textstyle\frac{4\pi}{M\Lambda^3}}\right)$\\\hline$2C_{4;ss}+\tilde{C}_{4;ss}($MeV$^{-6}$)&$-4.27\times10^{-12}$&$10^{-5}{\mathcal{O}}\left({\textstyle\frac{4\pi} {M\mu^3\Lambda^2}}\right)$&$-1.04\times10^{-13}$&$0.39\mathcal{O}\left({\textstyle\frac{4\pi}{M\Lambda^5}}\right)$\\\hline\hline\end{tabular}\end{center}\end{table}
\begin{figure}[t]\begin{center}
\resizebox{6.7cm}{!}{\includegraphics{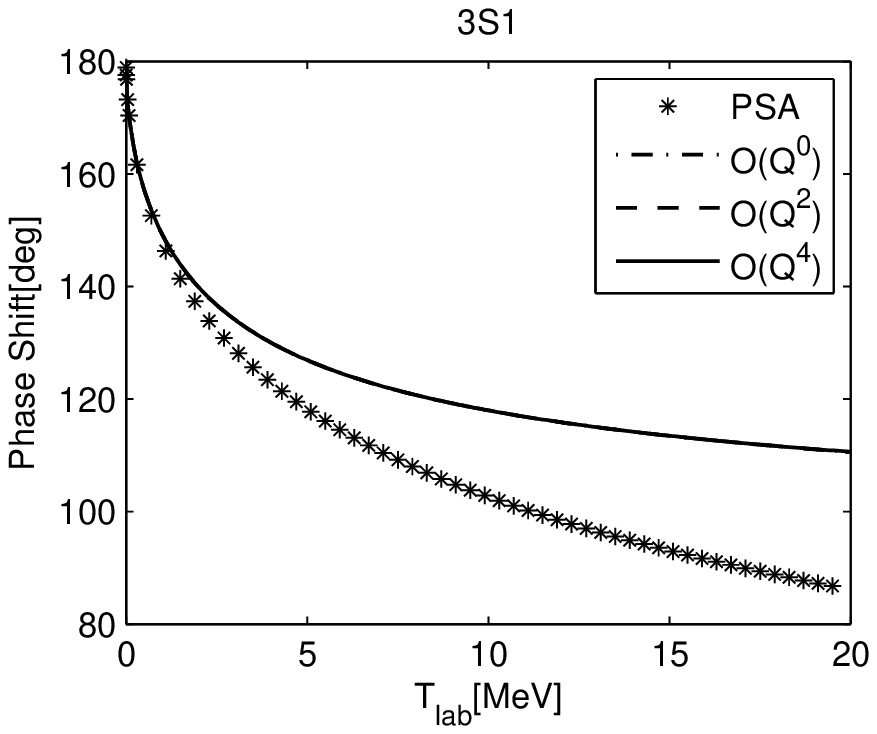}}\resizebox{6.7cm}{!}{\includegraphics{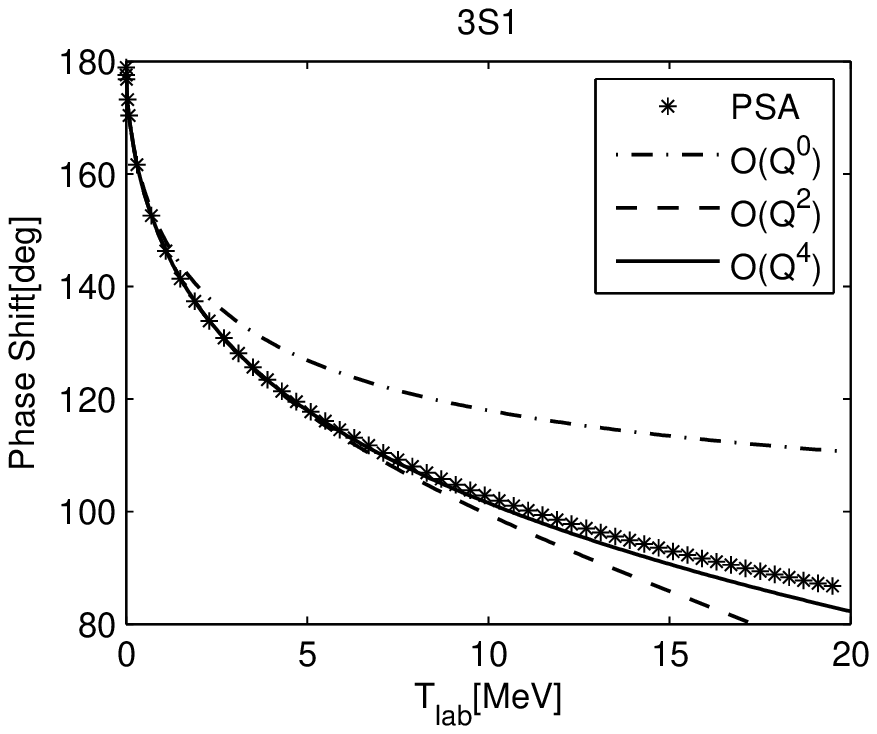}}
\caption{\small EFT($\not\!\!\pi$) predictions and PSA data of the $^3S_1$ phase shifts with fitting interval of laboratory energy $\texttt{T}_{\texttt{\tiny lab}} \in(0,3]$ MeV. (a) Left: Scenario B ($J_0\propto35$ MeV); (b) Right: Scenario C ($J_0\propto138$ MeV).}\label{PSfig}\end{center}\end{figure}

The predictions of the phase shifts over the range $\texttt{T}_{\texttt{\tiny lab}}\in[3,15]$ MeV are demonstrated in Fig.~\ref{PSfig}, from which we could see that: (1) The predictions in each scenario are improved systematically as truncation order increases, a natural merit of EFT description; (2) The scenario C predictions are closer to the PSA curve than scenario B as truncation order increases.
\section{Discussions and summary}\label{sec:4}Here, we wish to remark on the various approaches proposed and/or adopted in literature in the light of EFT scenario discussed so far. In many papers, a finite cut-off of various sorts are used to remove the divergences, whose rationality could be seen as below: The finite cut-offs essentially play the roles of the complementary parameters, which implement loop level subtractions effectively. As noted in Section~\ref{sec:2.4}, an EFT description usually breaks down at scales much higher than the upper scale of EFT\cite{EpelM,EpGe}. Then, such cut-offs must be judiciously incorporated to play the roles of a physical upper scale, otherwise, it would make no field-theoretical sense. While in the perturbation like treatments\cite{BBSvK,NTvK,LvK,BKV}, the complementary parameters are also incorporated in various disguises: They are either introduced as separation scales or 'allocated' somehow in the couplings according to certain modified EFT power counting. In such approaches, convergence becomes an issue. Similarly, in the approaches adopted in Refs.\cite{FTT,YEP1,YEP2,YEP3}, loop level subtractions are also effectively performed and complementary parameters also show up (ones that could not be readily absorbed into couplings). The scenario structures might also be realized somehow by incorporating unconventional degrees and the associated couplings\cite{soto-dibaryon1,soto-dibaryon2}. In a sense, the various approaches in literature seem to 'converge' to the EFT scenario explicated in this report. In the contexts beyond few-body systems, it was noted in Ref.\cite{EPVRAM} that the perturbative like treatments of pion exchanges still has some problems to fix, while the other main choice seems more efficient\cite{review2a,review2b,review3,review4,review5,bookNN}.

Despite being more sophisticated in structures, the pionfull theory essentially face the same obstacles in nonperturbative regime: Not all divergences in LSE could be absorbed by the couplings available at a given order of EFT truncation. Again this mismatch means that some parameters from convolution have to be separately determined through physical boundaries. In short of closed-form $T$ matrices for pionfull theory, it might be instructive to study what could be inferred from the notion of EFT scenario delineated above for the treatment of pionfull theory. Phenomenological descriptions of phase shifts and mixing angles of various channels of $NN$ scattering using our closed-form $T$ matrices will be given in a separate report\cite{Phaseshifts}. Nonperturbative running couplings at higher orders and relations between closed-form and 'perturbative' $T$ matrices will be studied elsewhere\cite{ptnpt}. More applications of our approach and the scenario notion within and beyond nucleon systems will also be pursued in the future.

In summary, the closed-form $T$ matrices for $NN$ scattering in the coupled channels $^3S_1$$-$$^3D_1$ were presented and explored in a general parametrization of divergent integrals within the realm of EFT($\not\!\!\pi$), leading us to the following findings: Intrinsic mismatches exist between the EFT couplings and the finitely many nonperturbative divergences involved, subtractions must be performed at loop level with the unmatched parameters turned into physical ones. Several typical scenarios were then examined and analyzed in terms of effective range expansion and the scenarios with unconventional couplings seem to be pathological and also disfavored by PSA data, in contrast to a simple scenario with conventional couplings. This status is also supported by the preliminary EFT($\not\!\pi$) predictions of $^3S_1$ phase shifts. The utilities of fine tuning are demonstrated in several places and naturally interpreted in the underlying theory perspective. The various approaches in the literature on $NN$ scattering were also addressed in light of EFT scenario. Our investigation has been performed in a general way that is applicable to any consistent EFT dominated by contact or short-distance interactions.
\section*{Acknowledgments}
The project is supported in part by the Kavli Institute of Theoretical Physics China and by the Ministry of Education of China. The author is grateful to Bira Van Kolck, Yu-Qi Chen, Yu Jia, Fan Wang, Antonio Pineda, Hong-Ying Jin, Dao-Neng Gao and other participants at the KITPC program '{\em EFT's in Particle and Nuclear Physics}' for many helpful and enlightening conversations over the EFT topics. The author is also grateful to an anonymous referee for his/her criticisms and suggestions that greatly
improved the presentation of the manuscript.
\appendix
\section{}Suppose $X_{AB} (A,B=1,2)$ are four $n\times n$ matrices, then the super matrix\bea&&\underline{X}\equiv\left(\begin{array}{cc}X_{11}&X_{12}\\X_{21}&X_{22} \end{array}\right)\Rightarrow\underline{X}^{-1}=\displaystyle\left(\begin{array}{cc}\left(\underline{X}^{-1}\right)_{11},&\left(\underline{X}^{-1}\right)_{12}\\\left( \underline{X}^{-1}\right)_{21},&\left(\underline{X}^{-1}\right)_{22}\end{array}\right),\nonumber\\&&\left(\underline{X}^{-1}\right)_{11}=(X_{11}-X_{12}X^{-1}_{22}X_{21}) ^{-1},\ \left(\underline{X}^{-1}\right)_{12}=(X_{21}-X_{22}X_{12}^{-1}X_{11})^{-1},\nonumber\\&&\left(\underline{X}^{-1}\right)_{21}=(X_{12}-X_{11}X_{21}^{-1}X_{22}) ^{-1},\ \left(\underline{X}^{-1}\right)_{22}=(X_{22}-X_{21} X_{11}^{-1}X_{12})^{-1}.\eea When $X_{AB}$ ($A\neq B$) are singular, we have,\bea&&\left(\underline{X}^{-1} \right)_{12}=(X_{12}X_{22}^{-1}X_{21}-X_{11})^{-1}X_{12}X_{22}^{-1},\ \left(\underline{X}^{-1}\right)_{21}=X_{22}^{-1}X_{21}(X_{12}X_{22}^{-1}X_{21}-X_{11})^{-1},\\&& \left(\underline{X}^{-1}\right)_{12}=X_{11}^{-1}X_{12}(X_{21}X_{11}^{-1}X_{12}-X_{22})^{-1},\ \left(\underline{X}^{-1}\right)_{21}=(X_{22}-X_{21}X_{11}^{-1}X_{12})^{-1} X_{21}X_{11}^{-1}.\eea If three of the sub matrices are singular, then $\underline{X}^{-1}$ does not exist at all.

For $1-\underline{\lambda}\underline{{\mathcal{I}}}(E)$ in Section~\ref{sec:1}, we have$$\left(1-\underline{\lambda}\underline{{\mathcal{I}}}(E)\right)^{-1}= \displaystyle\left(\begin{array}{cc}\tilde{\mathcal{K}}_{ss},&\tilde{\mathcal{K}}_{ss}\lambda_{sd}{\mathcal{I}}\mathcal{K}_{dd}\\\tilde{\mathcal{K}}_{dd}\lambda_{ds} {\mathcal{I}}\mathcal{K}_{ss},&\tilde{\mathcal{K}}_{dd}\\\end{array}\right),$$ with $$\mathcal{K}_{xx}\equiv(1-\lambda_{xx}{\mathcal{I}})^{-1},\ \tilde{\mathcal{K}}_{xx} \equiv(1-\tilde{\lambda}_{xx}{\mathcal{I}})^{-1},$$ where $\tilde{\lambda}_{xx}$ is defined in Eq.(\ref{tildelambda1}) and $x=s,d$.
\section{}At order $\Delta=4$, we have\bea\Delta U_1\equiv\left(\begin{array}{ccc}0&1&p^2\\1&p^2&p^4\\p^2&p^4&p^6\\\end{array}\right),\ \Delta U_2\equiv\left( \begin{array}{ccc}0&0&1\\0&1&p^2\\1&p^2&p^4\\\end{array}\right),\ \Delta U_3\equiv\left(\begin{array}{ccc}0&0&0\\0&0&1\\0&1&p^2\\\end{array}\right),\ \Delta U_4\equiv \left(\begin{array}{ccc}0&0&0\\0&0&0\\0&0& 1\\\end{array}\right).\nonumber\eea
\section{}Introducing the following parametrization of $[{\mathcal{N}}_{\cdots},{\mathcal{D}}_{\cdots}]$: $${\mathcal{N}}_0\equiv\sum_{j=0}^3\nu_{0;j}p^{2j},\ {\mathcal {N}}_1\equiv\sum_{j=0}^2\nu_{1;j}p^{2j},\ {\mathcal{D}}_0\equiv\sum_{j=0}^4\delta_{0;j}p^{2j},\ {\mathcal{D}}_1\equiv\sum_{j=0}^3\delta_{1;j}p^{2j},\ {\mathcal{D}}_{sd} \equiv\sum_{j=0}^2\delta_{sd;j}p^{2j},$$ the concrete expressions of the coefficients $[\nu_{\cdots},\delta_{\cdots}]$ read as follows,\bea\nu_{0;0}=&&(1-C_{2;ss}J_3-C_ {4;ss}J_5)^2-(\tilde{C}_{4;ss}+C_{4;dd})(1-C_{4;ss}J_5)^2J_5-C_{0;ss}\tilde{C}_{4;ss}J_3^2\nonumber\\&&+2(1-C_{4;ss}J_5)[(C_{2;ss}C_{4;dd}-C_{2;sd}\tilde{C}_{4;sd})J_5 -(C_{4;ss}\tilde{C}_{4;ss}+C_{4;sd}\tilde{C}_{4;sd})J_7]J_3\nonumber\\&&+[(2C_{2;ss}C_{2;sd}\tilde{C}_{4;sd}-C_{2;ss}^2C_{4;dd}-C_{2;sd}^2\tilde{C}_{4;ss})J_5+2C_{4;sd} (C_{2;ss}\tilde{C}_{4;sd}\nonumber\\&&-C_{2;sd}\tilde{C}_{4;ss})J_7-\tilde{C}_{4;ss}(C_{4;ss}^2+C_{4;sd}^2)J_9]J_3^2+(\tilde{C}_{4;ss}C_{4;dd}-\tilde{C}_{4;sd}^2)\{C_ {0;ss}J_3^2J_5\nonumber\\&&+(1-C_{4;ss}J_5)^2J_5^2+2C_{4;ss}J_3J_5J_7(1-C_{4;ss}J_5)+C_{4;sd}^2J_3^2(J_5J_9-J_7^2)\nonumber\\&&+C_{4;ss}^2J_3^2J_5J_9\};\\\nu_{0;1}=&&\{ -\tilde{C}_{4;ss}(1-C_{4;ss}J_5)^2-C_{4;dd}(1-C_{2;ss}J_3)^2-2(C_{4;ss}+C_{2;sd}\tilde{C}_{4;sd}J_3)\nonumber\\&&\times(1-C_{2;ss}J_3)+2(C_{4;ss}^2+2C_{4;ss}C_{4;dd}- C_{4;sd}\tilde{C}_{4;sd})J_5-C_{2;sd}^2\tilde{C}_{4;ss}J_3^2\nonumber\\&&+2[(C_{2;ss}\tilde{C}_{4;sd}-C_{2;sd}\tilde{C}_{4;ss})C_{4;sd}+2(C_{2;sd}\tilde{C}_{4;sd}-C_ {2;ss}C_{4;dd})C_{4;ss}]J_3J_5\nonumber\\&&-3C_{4;dd}C_{4;ss}^2J_5^2+2C_{4;ss}C_{4;sd}\tilde{C}_{4;sd}(J_5^2+J_3J_7)+\tilde{C}_{4;ss}(C_{4;ss}^2-C_{4;sd}^2)J_3J_7 \nonumber\\&&+(\tilde{C}_{4;ss}C_{4;dd}-\tilde{C}_{4;sd}^2)[C_{0;ss}J_3^2+2(1-C_{4;ss}J_5)^2J_5+2C_{4;ss}J_3J_7+C_{4;ss}^2J_3\nonumber\\&&\times(J_3J_9-3J_5J_7)+C_{4;sd} ^2(J_3J_9-J_5J_7)J_3]\}J_3;\\\nu_{0;2}=&&\{C_{4;ss}^2+(C_{4;ss}C_{4;dd}-C_{4;sd}\tilde{C}_{4;sd})(2-2C_{2;ss}J_3-3C_{4;ss}J_5)+(C_{4;ss}\tilde{C}_{4;sd}\nonumber\\&&-C_ {4;sd}\tilde{C}_{4;ss})(2C_{2;sd}J_3+C_{4;sd}J_5)+(\tilde{C}_{4;ss}C_{4;dd}-\tilde{C}_{4;sd}^2)[(1-C_{4;ss}J_5)^2\nonumber\\&&-(C_{4;ss}^2+C_{4;sd}^2)J_3J_7]\}J_3^2;\\ \nu_{0;3}=&&(2C_{4;ss}C_{4;sd}\tilde{C}_{4;sd}-C_{4;ss}^2C_{4;dd}-\tilde{C}_{4;ss}C_{4;sd}^2)J_3^3;\\\nu_{1;0}=&&C_{4;dd}(1-C_{2;ss}J_3-C_{4;ss}J_5)^2+2C_{2;sd}\tilde{C} _{4;sd}(1-C_{4;ss}J_5)J_3+(C_{2;sd}^2\tilde{C}_{4;ss}\nonumber\\&&-2C_{2;ss}C_{2;sd}\tilde{C}_{4;sd})J_3^2+(\tilde{C}_{4;sd}^2-C_{4;dd}\tilde{C}_{4;ss})[C_{0;ss}J_3^2+ (1-C_{4;ss}J_5)^2J_5\nonumber\\&&+2C_{4;ss}(1-C_{4;ss}J_5)J_3J_7+(C_{4;ss}^2+C_{4;sd}^2)J_3^2J_9];\\\nu_{1;1}=&&\{2(C_{4;sd}\tilde{C}_{4;sd}-C_{4;ss}C_{4;dd})(1-C_{2;ss} J_3-C_{4;ss}J_5)+2C_{2;sd}(C_{4;sd}\tilde{C}_{4;ss}\nonumber\\&&-C_{4;ss}\tilde{C}_{4;sd})J_3+(\tilde{C}_{4;sd}^2-C_{4;dd}\tilde{C}_{4;ss})[(1-J_5C_{4;ss})^2+(C_{4;ss}^2 +C_{4;sd}^2)\nonumber\\&&\times J_3J_7]\}J_3;\\\nu_{1;2}=&&(C_{4;ss}^2C_{4;dd}+\tilde{C}_{4;ss}C_{4;sd}^2-2C_{4;ss}C_{4;sd}\tilde{C}_{4;sd})J_3^2;\\\delta_{0;0}=&&[C_ {0;ss}+(C_{4;ss}^2+C_{4;sd}^2)J_9][1-(\tilde{C}_{4;ss}+C_{4;dd})J_5]+(C_{2;ss}^2+C_{2;sd}^2)J_5\nonumber\\&&+2(C_{2;ss}C_{4;ss}+C_{2;sd}C_{4;sd})J_7+(C_{2;sd}\tilde{C}_ {4;sd}-C_{2;ss}C_{4;dd})(C_{2;ss}J_5^2\nonumber\\&&+2C_{4;ss}J_5J_7)+(C_{2;ss}\tilde{C}_{4;sd}-C_{2;sd}\tilde{C}_{4;ss})(C_{2;sd}J_5^2+2C_{4;sd}J_5J_7)\nonumber\\&&+(C_ {4;ss}^2\tilde{C}_{4;ss}+2C_{4;ss}C_{4;sd}\tilde{C}_{4;sd}+C_{4;sd}^2C_{4;dd})J_7^2+(\tilde{C}_{4;ss}C_{4;dd}-\tilde{C}_{4;sd}^2)\nonumber\\&&\times[C_{0;ss}J_5^2+(C_ {4;ss}^2+C_{4;sd}^2)(J_5^2J_9-J_5J_7^2)];\\\delta_{0;1}=&&2C_{2;ss}+(C_{2;sd}^2-C_{2;ss}^2)J_3+(\tilde{C}_{4;ss}-C_{4;dd})[C_{0;ss}J_3+(C_{4;ss}^2+C_{4;sd}^2)J_3J_9 \nonumber\\&&-C_{4;sd}^2J_5J_7]+2C_{2;sd}C_{4;sd}J_5+2(C_{2;sd}\tilde{C}_{4;sd}-C_{2;ss}C_{4;dd})[J_5+C_{4;ss}J_3J_7]\nonumber\\&&+(C_{4;ss}^2+C_{4;sd}^2+2C_{4;ss}\tilde {C}_{4;ss}+2C_{4;sd}\tilde{C}_{4;sd})J_7+2(C_{2;ss}\tilde{C}_{4;sd}-C_{2;sd}\tilde{C}_{4;ss})\nonumber\\&&\times C_{4;sd}(J_5^2-J_3J_7)+[2C_{4;ss}C_{4;sd}\tilde{C}_ {4;sd}-C_{4;ss}^2(\tilde{C}_{4;ss}+C_{4;dd})]J_5J_7+(\tilde{C}_{4;sd}^2\nonumber\\&&-\tilde{C}_{4;ss}C_{4;dd})[2C_{4;ss}J_5+(C_{4;ss}^2-C_{4;sd}^2)(J_3J_7-J_5^2)]J_7; \eea\bea\delta_{0;2}=&&2C_{4;ss}+\tilde{C}_{4;ss}+(C_{4;ss}+C_{4;dd})[(1-C_{4;ss}J_5)^2-1-2C_{2;ss}J_3+(C_{4;sd}^2\nonumber\\&&-C_{4;ss}^2)J_3J_7]+2(C_{4;sd}+\tilde{C}_ {4;sd})C_{2;sd}J_3+[C_{4;sd}^2-C_{4;ss}^2+2C_{4;sd}\tilde{C}_{4;sd})]J_5\nonumber\\&&+C_{2;sd}(C_{2;sd}\tilde{C}_{4;ss}-C_{2;ss}\tilde{C}_{4;sd})J_3^2+(C_{2;ss}C_{4;dd} -C_{2;sd}\tilde{C}_{4;sd})(C_{2;ss}J_3^2\nonumber\\&&+2C_{4;ss}J_3J_5)-C_{4;sd}^2\tilde{C}_{4;ss}J_5^2+(\tilde{C}_{4;sd}^2-\tilde{C}_{4;ss}C_{4;dd})\{(1-C_{4;ss}J_5)^2 J_5\nonumber\\&&+2(1-C_{4;ss}J_5)C_{4;ss}J_3J_7+[C_{0;ss}+(C_{4;ss}^2+C_{4;sd}^2)J_9]J_3^2\};\\\delta_{0;3}=&&\{C_{4;sd}^2-C_{4;ss}^2+2(C_{4;ss}C_{4;dd}+C_{4;sd}\tilde {C}_{4;sd})+2(C_{4;ss}C_{4;dd}-C_{4;sd}\tilde{C}_{4;sd})\nonumber\\&&\times(C_{2;ss}J_3+C_{4;ss}J_5)+2(C_{4;sd}\tilde{C}_{4;ss}-C_{4;ss}\tilde{C}_{4;sd})C_{2;sd}J_3- (\tilde{C}_{4;ss}C_{4;dd}\nonumber\\&&-\tilde{C}_{4;sd}^2)[(1-C_{4;ss}J_5)^2-(C_{4;ss}^2+C_{4;sd}^2)J_3J_7]\}J_3;\\\delta_{0;4}=&&(C_{4;ss}^2C_{4;dd}+\tilde{C}_{4;ss}C_ {4;sd}^2-2C_{4;ss}C_{4;sd}\tilde{C}_{4;sd})J_3^2;\\\delta_{1;0}=&&[C_{0;ss}+(C_{4;ss}^2+C_{4;sd}^2)J_9]C_{4;dd}-C_{2;sd}^2+[C_{2;ss}^2C_{4;dd}+C_{2;sd}^2\tilde{C}_{4;ss} -2C_{2;ss}\nonumber\\&&\times C_{2;sd}\tilde{C}_{4;sd}]J_5+2(C_{2;ss}C_{4;dd}-C_{2;sd}\tilde{C}_{4;sd})C_{4;ss}J_7+(\tilde{C}_{4;ss}C_{4;dd}-\tilde{C}_{4;sd}^2) \nonumber\\&&\times[C_{4;ss}^2J_7^2-C_{0;ss}J_5-(C_{4;ss}^2+C_{4;sd}^2)J_5J_9];\\\delta_{1;1}=&&2[C_{2;ss}C_{4;dd}-C_{2;sd}(C_{4;sd}+\tilde{C}_{4;sd})]+(\tilde{C}_{4;ss} C_{4;dd}-\tilde{C}_{4;sd}^2)[C_{0;ss}J_3+2C_{4;ss}\nonumber\\&&\times J_7+(C_{4;ss}^2+C_{4;sd}^2)J_3J_9+(C_{4;sd}^2-C_{4;ss}^2)J_5J_7]+(C_{2;sd}\tilde{C}_{4;sd}-C_{2;ss} C_{4;dd})\nonumber\\&&\times C_{2;ss}J_3+(C_{2;sd}\tilde{C}_{4;ss}-C_{2;ss}\tilde{C}_{4;sd})(2C_{4;sd}J_5 -C_{2;sd}J_3)+[(C_{4;ss}^2-C_{4;sd}^2)\nonumber\\&&\times C_ {4;dd}-2C_{4;ss}C_{4;sd}\tilde{C}_{4;sd}]J_7;\\\delta_{1;2}=&&(\tilde{C}_{4;ss}C_{4;dd}-\tilde{C}_{4;sd}^2)[(1-C_{4;ss}J_5)^2-(C_{4;ss}^2+C_{4;sd}^2)J_3J_7]+2(C_{4;ss}C _{4;dd}\nonumber\\&&-C_{4;sd}\tilde{C}_{4;sd})(1-C_{2;ss}J_3)-C_{4;sd}^2+2(C_{4;ss}\tilde{C}_{4;sd}-C_{4;sd}\tilde{C}_{4;ss})C_{2;sd}J_3\nonumber\\&&+(\tilde{C}_{4;ss}C _{4;sd}^2-C_{4;ss}^2C_{4;dd})J_5;\\\delta_{1;3}=&&-(C_{4;ss}^2C_{4;dd}+\tilde{C}_{4;ss}C_{4;sd}^2-2C_{4;ss}C_{4;sd}\tilde{C}_{4;sd})J_3;\\\delta_{sd;0}=&&C_{2;sd}(1-C_ {2;ss}J_3-C_{4;ss}J_5)+[C_{0;ss}+(C_{4;ss}^2+C_{4;sd}^2)J_9]\tilde{C}_{4;sd}J_3+(C_{4;ss}\nonumber\\&&\times\tilde{C}_{4;sd}+C_{4;sd}C_{4;dd})(1-C_{4;ss}J_5)J_7+(C_ {2;sd}\tilde{C}_{4;sd}-C_{2;ss}C_{4;dd})C_{4;sd}J_3J_7\nonumber\\&&+[C_{2;ss}\tilde{C}_{4;sd}-C_{2;sd}\tilde{C}_{4;ss}+(\tilde{C}_{4;sd}^2-C_{4;dd}\tilde{C}_{4;ss})C_ {4;sd}J_7][(1-C_{4;ss}J_5)J_5\nonumber\\&&+C_{4;ss}J_3J_7];\\\delta_{sd;1}=&&C_{4;sd}+\tilde{C}_{4;sd}-[C_{2;ss}C_{4;sd}+C_{2;sd}C_{4;ss}+(\tilde{C}_{4;sd}+C_{4;dd})C_ {4;ss}C_{4;sd}J_7]J_3\nonumber\\&&-[C_{4;ss}C_{4;sd}+\tilde{C}_{4;ss}C_{4;sd}(1-C_{4;ss}J_5)+C_{4;ss}\tilde{C}_ {4;sd}]J_5;\\\delta_{sd;2}=&&-C_{4;ss}C_{4;sd}J_3;\eea and \bea\nu_{0;3}+\nu_{1;2}J_3=0,\ \delta_{0;4}+\delta_{1;3}J_3=0,\ \delta_{0;4}= \nu_{1;2},\ \cdots.\eea
\section{}In this appendix, we present the rigorous proof of the $\mathcal{I}_0$-independence of the fractional parts of $\bf{T}^{-1}$ in $^3S_1$$-$$^3D_1$. To this end, we need to prove the unitarity of $\textbf{T}$ first.

Using the super matrix notations introduced in Sect.~\ref{sec:1.1}, the algebraic LSE's and their conjugates read,\bea\label{aLSE}&&\underline{\tau}=\underline{\lambda} +\underline{\lambda}\underline{{\mathcal{I}}}\underline{\tau},\\\label{aLSEC}&&\underline{\tau}^{\dagger}=\underline{\lambda}+\underline{\tau}^{\dagger}\underline {{\mathcal{I}}}^{\dagger}\underline{\lambda},\eea where the transpose symmetry of $\underline {\lambda}$ has been used. Now, multiplying Eq.(\ref{aLSE}) from the right by $\underline{{\mathcal{I}}}\underline{\tau}^{\dagger}$ and multiplying Eq.(\ref{aLSEC})from the left by $\underline{\tau}\underline{{\mathcal{I}}}^{\dagger}$, we could find that,\bea&&\underline{\tau}\underline{{\mathcal{I}}}\underline{\tau}^{\dagger}=\underline{\lambda}\underline{{\mathcal{I}}}\underline{\tau}^{\dagger}+\underline {\lambda}\underline{{\mathcal{I}}}\underline{\tau}\underline{{\mathcal{I}}}\underline{\tau}^{\dagger}\Rightarrow\underline{\tau}\underline{{\mathcal{I}}}\underline {\tau} ^{\dagger}=-\underline{\tau}^{\dagger}+\left(1-\underline{\lambda}\underline{{\mathcal{I}}}\right)^{-1}\underline{\tau}^{\dagger},\\&&\underline{\tau}\underline{ {\mathcal{I}}}\underline{\tau}^{\dagger}=\underline{\tau}\underline{{\mathcal{I}}}\underline{\lambda}+\underline{\tau}\underline{{\mathcal{I}}}\underline{\tau}^{\dagger} \underline{{\mathcal{I}}}^{\dagger}\underline{\lambda}\Rightarrow\underline{\tau}\underline{{\mathcal{I}}}^{\dagger}\underline{\tau}^{\dagger}=-\underline{\tau}+ \underline{\tau}(1-\underline{{\mathcal{I}}}^{\dagger}\underline{\lambda})^{-1}.\eea Noting that\bea\left(1-\underline{\lambda}\underline{{\mathcal{I}}}\right)^{-1} \underline{\tau}^{\dagger}=\left(1 -\underline{\lambda}\underline{{\mathcal{I}}}\right)^{-1}\lambda(1-\underline{{\mathcal{I}}}^{\dagger}\underline{\lambda})^{-1}= \underline{\tau}(1-\underline{{\mathcal{I}}}^{\dagger}\underline{\lambda})^{-1},\eea we finally have\bea\underline{\tau}(\underline{{\mathcal{I}}}^{\dagger}-\underline {{\mathcal{I}}})\underline{\tau}^{\dagger}=\underline{\tau}^{\dagger}-\underline{\tau}.\eea This is the unitarity relation in terms of $\underline{\tau}$.

Now sandwiching Eq.(D.6) between $$\left(\begin{array}{cc}U&0\\0&U\end{array}\right)^T\ \text{and}\ \left(\begin{array}{cc}U&0\\0&U\end{array}\right),$$ we find\bea&& \left(\begin{array}{cc}U^T\tau_{ss}U&U^T\tau_{sd}U\\U^T\tau_{ds}U&U^T\tau_{dd}U\\\end{array}\right)^{\dagger}-\left(\begin{array}{cc}U^T\tau_{ss}U&U^T\tau_{sd}U\\U^T \tau_{ds}U&U^T\tau_{dd}U\\\end{array}\right)={\textbf{T}}^{\dagger}-{\textbf{T}}\nonumber\\&&=i\frac{Mp}{2\pi}\left(\begin{array}{cc}U^T\tau_{ss}&U^T\tau_{sd}\\U^T\tau _{ds}&U^T\tau_{dd}\\\end{array}\right)\left(\begin{array}{cc}UU^T&\bf{0}\\\bf{0}&UU^T\\\end{array}\right)\left(\begin{array}{cc}U^T\tau_{ss}&U^T\tau_{sd}\\U^T\tau_{ds}& U^T\tau_{dd}\\\end{array}\right)^{\dagger}\nonumber\\&&=i\frac{Mp}{2\pi}\left(\begin{array}{cc}U^T\tau_{ss}U&U^T\tau_{sd}U\\U^T\tau_{ds}U&U^T\tau_{dd}U\\\end{array} \right)\left(\begin{array}{cc}U^T\tau_{ss}U&U^T\tau_{sd}U\\U^T\tau_{ds}U&U^T\tau_{dd}U\\\end{array}\right)^{\dagger}=i\frac{Mp}{2\pi}{\textbf{T}}{\textbf{T}}^{\dagger}. \eea Then, we arrive at the unitarity \bea{\textbf{T}}^{-1}-({\textbf{T}}^{\dagger})^{-1}=i\frac{Mp}{2\pi}\textbf{I},\eea as claimed in Section~\ref{sec:1.3}.

Since $J_0$ always go with $i\frac{Mp}{4\pi}$, we conclude from the unitarity that\bea{\textbf{T}}^{-1}={\mathcal{I}}_0\textbf{I}+\Delta\textbf{R},\eea with $\Delta \textbf{R}$ being ${\mathcal{I}}_0$-independent. In fact, each element of $\Delta\textbf{R}$ takes the form of $\frac{\widetilde{\mathcal{N}}}{\widetilde{\mathcal{D}}}$ with $\widetilde{\mathcal{N}}$ and $\widetilde{\mathcal{D}}$ being polynomials in terms of $[C_{\cdots}],[J_{2m+1}]$ and $p^2$. The proof goes as below: It is easy to see that $\textbf{T}$ is a matrix made of rational functions in terms of $[C_{\cdots}],[J_{2m+1}]$ and $p^2$, thus $\textbf{T}^{-1}$ must also be such kind of matrix. As the only complex parameter ${\mathcal{I}}_0$ has been isolated, the rest must be a real rational matrix in terms of $[C_{\cdots}],[J_{2m+1}]$ and $p^2$. {\em Q.E.D.}

\end{document}